\documentclass[12pt]{article}
\pdfoutput=1
\usepackage{lmodern}

\usepackage{amsfonts,amsmath,amssymb,amstext,mathrsfs,epsfig,cite,setspace,bigstrut,framed}
\usepackage[all]{xy}
\usepackage{color,xcolor,soul}
\usepackage{bbm}
\usepackage{dsfont}
\usepackage{pifont}
\usepackage{latexsym}
\usepackage{graphics,graphicx} 
\usepackage[colorlinks = true, allcolors = blue!30!black!80!, linktocpage=true]{hyperref}
\usepackage[bbgreekl]{mathbbol}
\usepackage{fancyhdr}
\usepackage{lastpage}
\usepackage{etoolbox}
\usepackage{setspace}
\usepackage{titlesec}
\usepackage{cancel}
\usepackage{comment}
\usepackage[framemethod=tikz]{mdframed}
\usepackage{empheq}
\usepackage{tikz}
\usepackage{tcolorbox}
\tcbuselibrary{theorems}
\usepackage[font=footnotesize,labelfont=bf,justification=centerlast,width=.94\textwidth]{caption}

\def\k{\kappa}
\def\r{\rho}
\def\a{\alpha}

\def\b{\beta}

\def\g{\gamma}
\def\G{\Gamma}
\def\d{\delta}

\def\e{\epsilon}

\def\p{\pi}

\def\c{\chi}

\def\th{\theta}

\def\m{\mu}
\def\n{\nu}

\def\l{\lambda}
\def\L{\Lambda}
\def\s{\sigma}

\def\cV{{\cal V}}

\def\IR{\relax{\rm I\kern-.18em R}}

\def \z { {\bar z} }

\def\diag{{\rm diag}}

\def\IR{\relax{\rm I\kern-.18em R}}
\def\IL{\relax{\rm I\kern-.18em L}}

\def\inv{^{\raise.15ex\hbox{${\scriptscriptstyle -}$}\kern-.05em 1}}

\def\cV{{\cal V}}

\def\Tr{{\rm Tr}}

\makeatletter \@addtoreset{equation}{section} \makeatother
\renewcommand{\theequation}{\thesection.\arabic{equation}}

\def\del{\partial}

\def\){\right)}
\def\({\left( }
\def\]{\right] }
\def\[{\left[ }

\def\NO{\nonumber}

\newcommand{\be}{\begin{equation}}
\newcommand{\ee}{\end{equation}}

\def\bea{\begin{eqnarray}}
\def\eea{\end{eqnarray}}

\def\bal#1\eal{\begin{align}#1\end{align}}

\def\bald{\begin{aligned}}
\def\eald{\end{aligned}}

\def\bsub{\begin{subequations}}
\def\esub{\end{subequations}}

\def\beqx{\begin{displaymath}}
\def\eeqx{\end{displaymath}}

\newcommand{\bmat}{\left(\begin{array}}
\newcommand{\emat}{\end{array}\right)}

\def\a{\alpha}
\def\b{\beta}
\def\c{\chi}
\def\d{\delta}
\def\e{\epsilon}
\def\f{\phi}
\def\g{\gamma}
\def\h{\eta}
\def\j{\psi}
\def\k{\kappa}
\def\l{\lambda}
\def\m{\mu}
\def\n{\nu}
\def\o{\omega}
    
\def\p{\pi}

    \def\th{\theta}
\def\r{\rho}
\def\s{\sigma}
\def\t{\tau}
\def\x{\xi}
\def\z{\zeta}

\def\F{\Phi}
\def\G{\Gamma}

\def\L{\Lambda}
\def\O{\Omega}

\def\ve{\varepsilon}

\def\vf{\varphi}

\def\ca{{\cal A}}

\def\cc{{\cal C}}
\def\cd{{\cal D}}
\def\ce{{\cal E}}

\def\ci{{\cal I}}
\def\cj{{\cal J}}
\def\ck{{\cal K}}
\def\cl{{\cal L}}

\def\cn{{\cal N}}
\def\co{{\cal O}}
\def\cp{{\cal P}}
\def\cq{{\cal Q}}

\def\cs{{\cal S}}
\def\ct{{\cal T}}

\def\cv{{\cal V}}
\def\cw{{\cal W}}

\def\cz{{\cal Z}}

\def\bb#1{\ensuremath{\mathbb{#1}}} 

\def\bo{{\raise-.3ex\hbox{\large$\Box$}}}               
\def\pa{\partial}                                       
\def\face{{\raise.2ex\hbox{$\displaystyle \bigodot$}\mskip-2.2mu \llap {$\ddot
        \smile$}}}                                   
\def\>{\rangle}                                      
\def\<{\langle}                                      

\def\wt#1{\widetilde{#1}}                            
\def\lbar#1{\ensuremath{\overline{#1}}}              
\def\leftrightarrowfill{$\mathsurround=0pt \mathord\leftarrow \mkern-6mu
        \cleaders\hbox{$\mkern-2mu \mathord- \mkern-2mu$}\hfill
        \mkern-6mu \mathord\rightarrow$}        
\def\dvec#1{\vbox{\ialign{##\crcr
        \leftrightarrowfill\crcr\noalign{\kern-1pt\nointerlineskip}
        $\hfil\displaystyle{#1}\hfil$\crcr}}}           
\def\Tr{{\rm Tr \,}}                                    
\def\diag{{\rm diag \,}}                                

\def\-{\hphantom{-}}

\setul{0.5ex}{0.3ex}
    \definecolor{Red}{rgb}{1,0.0,0.0}
    \setulcolor{Red}

\newmdenv[%
middlelinecolor=gray!100!,
middlelinewidth=1.5pt,
backgroundcolor=gray!10!,
roundcorner=3pt
]{myidentity}

\mdfdefinestyle{mystyle}{%
middlelinecolor=gray!80!white,
middlelinewidth=1pt,
backgroundcolor=blue!5!white,
roundcorner=10pt,
subtitlebelowline=true,
frametitlefont={\normalfont\bfseries\sffamily\color{gray!90!white}},
innertopmargin=.1cm,
innerbottommargin=.3cm,
}

\def\bebx #1 \eebx{\begin{empheq}[box={\tcbhighmath[colframe=gray!90!white,colback=white]}]{align} #1 \end{empheq}}

\def\bbxd#1\ebxd{\begin{myidentity} \vskip -.4cm #1 \end{myidentity}\vskip-.2cm}

\def\bframe#1{\begin{mdframed}[style=mystyle,frametitle={#1}]}
\def\eframe{\end{mdframed}}

\begin{document}
\begin{titlepage}
\vfill
\begin{flushright}
{\normalsize APCTP Pre2024-002}\\
\end{flushright}
\vfill

\begin{center}
\vbox{\center\LARGE\bf  Supersymmetric Casimir energy on\\ $\cn=1$ conformal supergravity backgrounds}\vspace{5mm}

\vbox{\center Pantelis Panopoulos$^{a}$\thinspace,  
Ioannis Papadimitriou$^{b}$}\vspace{5mm}

\vbox{\center\em $^{a}$  Asia Pacific Center for Theoretical Physics, Postech, Pohang 37673, Korea}
\vbox{\center\em $^{b}$Division of Nuclear and Particle Physics, Department of Physics,
National and Kapodistrian University of Athens, GR-157 84 Athens, Greece}

\vbox{\center \href{mailto: Panopoulospant@gmail.com}{\tt Pantelis.Panopoulos@apctp.org}, 
\href{mailto: ioannis.papadimitriou@phys.uoa.gr}{\tt Ioannis.Papadimitriou@phys.uoa.gr}}\vspace{5mm}

\end{center}
\vfill

\begin{abstract}
We provide a first principles derivation of the supersymmetric Casimir energy of $\cn=1$ SCFTs in four dimensions using the supercharge algebra on general conformal supergravity backgrounds that admit Killing spinors. The superconformal Ward identities imply that there exists a continuous family of conserved R-currents on supersymmetric backgrounds, as well as a continuous family of conserved currents for each conformal Killing vector. These continuous families interpolate between the consistent and covariant R-current and energy-momentum tensor. The resulting Casimir energy, therefore, depends on two continuous parameters corresponding to the choice of conserved currents used to define the energy and R-charge. This ambiguity is in addition to any possible scheme dependence due to local terms in the effective action. As an application, we evaluate the general expression for the supersymmetric Casimir energy we obtain on a family of backgrounds with the cylinder topology $\bb R\times S^3$ and admitting a single Majorana supercharge. Our result is a direct consequence of the supersymmetry algebra, yet it resembles more known expressions for the non-supersymmetric Casimir energy on such backgrounds and differs from the supersymmetric Casimir energy obtained from the zero temperature limit of supersymmetric partition functions. We defer a thorough analysis of the relation between these results to future work.         

\end{abstract}

\vfill
\end{titlepage}

\tableofcontents
\addtocontents{toc}{\protect\setcounter{tocdepth}{3}}
\renewcommand{\thefootnote}{\arabic{footnote}}
\setcounter{footnote}{0}

\section{Introduction and summary of results}
\label{sec:intro}

The Casimir energy of a conformal field theory (CFT) is defined as its ground state energy. Placing the theory on the Lorentzian cylinder $\mathbb{R}\times S^{d-1}$ (Einstein universe), in the absence of other background fields, it is given by \cite{Birrell:1982ix}
\be\label{Casimir}
\ce_\text{Casimir}=\int_{S^{d-1}}d^{d-1}x\sqrt{-g}\,\< \ct_{tt}\>_{\text{g.s.}},
\ee
where $\ct_{\m\n}$ is the energy-momentum tensor and $\< ...\>_{\text{g.s.}}$ stands for the expectation value in the ground state of the CFT on the cylinder. This is determined by the conformal anomaly 
\be\label{trace}
\< \ct^\m_\m\>=\frac{1}{(4\p)^2}(a E-c W^2+k\square R)\,,
\ee
where $E$ is the Euler density, $W^2$ the square of the Weyl tensor, and the anomaly coefficients $a$ and $c$ depend on the specific CFT. The $k$-term is scheme-dependent and corresponds to the addition of a local counterterm proportional to $R^2$ in the effective action. 

To evaluate the Casimir energy, one may integrate the conformal anomaly \eqref{trace} in order to determine its contribution to the effective action, known as the Riegert action \cite{Riegert:1984kt,Fradkin:1983tg}. Its derivative with respect to the background metric computes the expectation value of the energy-momentum tensor in the ground state, and hence the Casimir energy \eqref{Casimir}. The Lorentzian computation can be found in \cite{Birrell:1982ix}, while the Euclidean in e.g. \cite{Herzog:2013ed,Assel:2015nca}. Notice that for time-independent backgrounds it is immaterial whether one evaluates the Casimir energy in the Lorentzian or Euclidean cylinder. The Lorentzian calculation involves an integral over a Cauchy surface, and can be related to the Euclidean one via Wick rotation. Consequently, the Casimir energy of a CFT is in general determined by the conformal anomaly coefficients $a$ and $c$, as well as any scheme-dependent terms, such as $k$. For example, the Casimir energy on the round unit sphere is given by \cite{Birrell:1982ix,Assel:2015nca}
\be\label{CasimirS0}
\ce_\text{Casimir}=\frac{3}{4}\Big(a-\frac k2\Big)\,.
\ee
The Casimir energy, therefore, is in general a scheme-dependent quantity. 

The procedure outlined above applies to any CFT, including supersymmetric conformal field theories (SCFTs). However, supersymmetry allows several alternative approaches to computing the supersymmetric Casimir energy \cite{Kim:2012ava,Assel:2014paa,Assel:2015nca,Bobev:2015kza,Brunner:2016nyk,BenettiGenolini:2016qwm,Papadimitriou:2017kzw}, while requiring that the renormalization scheme preserves supersymmetry reduces considerably the allowed scheme dependence. One common definition of the Casimir energy on (the Euclidean) $S^1_\b\times M_3$ with $M_3$ a three-manifold admitting supersymmetry is given by the ``zero temperature" limit, $\b\to\infty$, of the supersymmetric partition function, which isolates the ground state energy. Namely,
\be\label{CasPartition}
\ce^\text{susy}_\text{Casimir}=-\lim_{\b\to\infty}\frac{d}{d\b}\log Z^\text{susy}_{S^1_\b\times M_3}\,.
\ee

For $\cn=1$ SCFTs on $S_\b^1\times S^3_{b_1,b_2}$ with complex structure parameters $b_1$, $b_2$, it has been shown using supersymmetric localization that the partition function is proportional to the supersymmetric index, $\ci^\text{susy}_{S^1_\b\times S^3}$ \cite{Kim:2012ava,Assel:2014paa}, i.e.
\be\label{ZI}
Z^\text{susy}_{S^1_\b\times S^3}=e^{-\b \ce^\text{susy}_\text{Casimir}}\ci^\text{susy}_{S^1_\b\times S^3}\,,
\ee
where the exponent in the proportionality factor is given by 
\be\label{Esquashed}
\ce^\text{susy}_\text{Casimir}\left(b_1,b_2\right)=\frac{4\p}{3}(|b_1|+|b_2|)(a-c)+\frac{4\p}{27}\frac{(|b_1|+|b_2|)^3}{|b_1||b_2|}(3c-2a)\,,
\ee
and can be identified with the supersymmetric Casimir energy through the definition \eqref{CasPartition}. 
For the round $S^3$ metric with $b_1=b_2=\frac{\b}{2\p}$, the supersymmetric Casimir energy \eqref{Esquashed} becomes
\be\label{Esphere}
\ce_\text{Casimir}^\text{susy}\Big(\frac{\b}{2\p},\frac{\b}{2\p}\Big)=\frac{4}{27}(a+3c)\,.
\ee

Another approach to the evaluation of the supersymmetric Casimir energy was adopted in \cite{Assel:2015nca} and proceeds by dimensionally reducing an $\cn=1$ theory on $S_\b^1\times M_3$ to supersymmetric quantum mechanics, whose Hamiltonian can be expressed as the sum of chiral and Fermi multiplet Hamiltonians, namely
\be
\< H_\text{susy}\>=\sum_\text{chiral}\< H_\text{chiral}\>+\sum_\text{Fermi}\< H_\text{Fermi}\>\,.
\ee
The Casimir energy is then given by the ground state energy $\ce_\text{Casimir}=\langle H_\text{susy}\rangle$, which agrees with \eqref{Esquashed} and \eqref{Esphere}. Yet another approach was considered in\cite{Bobev:2015kza}, where it was conjectured that  the supersymmetric Casimir energy in $d$ even dimensions is given by an equivariant integral of the anomaly polynomial in two higher dimensions. 

The approach we follow in the preset work is based on the supersymmetry algebra. On a curved supersymmetric background, the supercharges satisfy (schematically) the algebra
\be\label{CasimirAlgDef}
\{Q^\dag,Q\}=H-Q_R-\ce_{\text{Casimir}}\,,
\ee
where $H$ is the Hamiltonian and $Q_R$ is the R-charge operator. Since in a supersymmetric vacuum $\<\{Q^\dag,Q\}\>=0$, it follows that
\be
\<H-Q_R\>=\ce_{\text{Casimir}}\,.
\ee
We determine the general form of the Casimir energy $\ce_{\text{Casimir}}$ in \eqref{CasimirAlgDef} by coupling a generic $\cn=1$ SCFT to background conformal supergravity and using the operator algebra obtained in \cite{Papadimitriou:2019gel}. This leads to the general expression for the supersymmetric Casimir energy given in eq.~\eqref{ECasimir} in terms of local curvatures and Killing spinor bilinears. Evaluating this general expression for the supersymmetric Casimir energy on backgrounds with the (Lorentzian) cylinder topology $\bb R\times S^3$ and admitting a single Majorana supercharge we arrive at the expression \eqref{ECasimir-S3}. This result is consistent with \eqref{CasimirS0}, but differs significantly from \eqref{Esquashed}. While our calculation is performed on Lorentzian backgrounds and the result \eqref{Esquashed} was obtained from Euclidean ones, this is not the reason for the discrepancy. The backgrounds involved in the two calculations are related by a Wick rotation, as discussed at length in  \cite{Cassani:2012ri,Cassani:2013dba}. Instead, we anticipate that the two results for the supersymmetric Casimir energy are related by a diffeomorphism-violating local counterterm that is implicitly added to the effective action when a holomorphic scheme is used, as is done in the evaluation of the partition function via supersymmetric localization \cite{BenettiGenolini:2016qwm,BenettiGenolini:2016tsn}. We defer a thorough analysis of the relation between the two results for the supersymmetric Casimir energy to a future publication \cite{SusyCasHol}.  

The rest of the paper is organized as follows. In section \ref{sec:WIDs} we review the $\cn=1$ superconformal Ward identities for the conformal current multiplet, including the superconformal anomalies. In section \ref{sec:BZ} we introduce the relevant Bardeen-Zumino terms for the R-current and energy-momentum tensor, which allow us to rewrite the superconformal Ward identities in terms of the covariant $R$-current and the energy-momentum tensor. In section \ref{sec:charges} we consider bosonic backgrounds with numerically zero anomalies, which include backgrounds that admit Killing spinors as a special case, and we show the existence of continuous families of conserved R-charges and charges associated with conformal Killing vectors. Section \ref{sec:SCE} is devoted to the detailed evaluation of the Casimir energy on such backgrounds using the anomalous transformation of the supercurrent obtained in \cite{Papadimitriou:2019gel}. This leads to the general BPS relation \eqref{genBPS} and the ``local charge'' \eqref{locCharge}. These are the main results of this work. As an illustrative example, in section \ref{sec:sphere}, we evaluate the Casimir energy on $\mathbb{R}\times S^3$, obtaining the expression \eqref{ECasimir-S3}.

\section{$\cn=1$ superconformal Ward identities}
\label{sec:WIDs}

The superconformal Ward identities that four-dimensional $\cn=1$ SCFTs satisfy can be determined by coupling the theory to $\cn=1$ off-shell conformal supergravity \cite{Kaku:1977pa,Kaku:1977rk,Kaku:1978nz,Townsend:1979ki}, whose field content consists of the vielbein $e^a_\m$, an Abelian gauge field $A_\m$, and a Majorana gravitino $\j_{\m}$, comprising 5+3 bosonic and 8 fermionic off-shell degrees of freedom. Treating the supergravity fields as external background fields, the SCFT partition function takes the form  
\be
\cz[e,A,\psi]=\int [\cd\Phi]e^{iS[\Phi,e,A,\psi]}\,,
\ee
where $\F$ denotes collectively all the microscopic degrees of freedom.\footnote{Evaluating the path integral over the microscopic degrees of freedom $\F$ in general requires the introduction of a regulator that explicitly breaks conformal symmetry. Although the regulated theory cannot couple consistently to conformal supergravity, the {\em renormalized} theory can, even in the presence of superconformal anomalies. Throughout this article we refer exclusively to the renormalized observables, which can consistently couple to conformal supergravity.} 

The logarithm of the partition function on an arbitrary supergravity background
\be\label{W}
\mathcal W[e,A,\psi]=-i\log\cz[e,A,\psi]\,,
\ee
is often referred to as the `quantum effective action' and amounts to the generating function of connected (renormalized) correlation functions of the current operators
\bbxd
\vskip.4cm
\be\label{consistent-currents}
\<\ct^\m_a\>=e^{-1}\frac{\d\mathcal W}{\d e^a_\m},\qquad \<\cj^\m\>=e^{-1}\frac{\d\mathcal W}{\d A_\m},\qquad \<\cs^\m\>=e^{-1}\frac{\d\mathcal W}{\d \bar\psi_\m}\,,
\ee
\ebxd
where $e\equiv\det(e^a_\m)$ and the notation $\<\cdot\>$ denotes one-point functions in the presence of arbitrary sources, i.e. on an arbitrary supergravity background. These operators comprise the conformal multiplet of currents and satisfy a set of superconformal Ward identities. 

$\cn=1$ conformal supergravity is a gauge theory of the $\cn=1$ superconformal algebra \cite{Kaku:1977pa,Kaku:1977rk,Kaku:1978nz,Townsend:1979ki} (see \cite{VanNieuwenhuizen:1981ae,deWit:1981vgr,deWit:1983qkc,Fradkin:1985am} and chapter 16 of \cite{Freedman:2012zz} for pedagogical reviews). Its local symmetries consist of diffeomorphisms $\x^\m(x)$, Weyl transformations $\s(x)$, local frame rotations $\l^{ab}(x)$, $U(1)$ R-symmetry transformations $\th(x)$, as well as $\cq$- and $\cs$-supersymmetry transformations, parameterized respectively by the local spinors $\ve(x)$ and $\h(x)$. Under these, the fields of $\cn=1$ conformal supergravity transform as   
\bal\label{sugra-trans}
\d e^a_\m=&\;\x^\l\pa_\l e^a_\m+e^a_\l\pa_\m\x^\l-\l^a{}_b e^b_\m+\s e^a_\m-\frac12\lbar\j_\m\g^a\ve\,,\NO\\
\d\j_\m=&\;\x^\l\pa_\l\j_\m+\j_\l\pa_\m\x^\l-\frac14\l_{ab}\g^{ab}\j_\m+\frac12\s\j_\m+\cd_\m\ve-\g_\m\h- i\g^5\th\j_\m\,,\NO\\
\d A_\m=&\;\x^\l\pa_\l A_\m+A_\l\pa_\m\x^\l+\frac{3i}{4}\lbar\f_\m\g^5\ve-\frac{3i}{4}\lbar\j_\m\g^5\h+\pa_\m\th\,,
\eal
where 
\be\label{phi}
\f_\m\equiv -\frac16\big(4\d^{[\r}_\m\d^{\s]}_\n+i\g^5 \e_{\m\n}{}^{\r\s}\big)\g^\n \cd_\r\j_\s\,,
\ee
and the spinor covariant derivatives are given by
\bal\label{covD}
\cd_\m\j_\n\equiv&\;\Big(\pa_\m+\frac14\o_\m{}^{ab}(e,\j)\g_{ab}+i\g^5A_\m\Big)\j_\n-\G^\r_{\m\n}\j_\r\equiv \big(\mathscr{D}_\m+i\g^5A_\m\big)\j_\n\,,\NO\\
\cd_\m\ve\equiv&\;\Big(\pa_\m+\frac14\o_\m{}^{ab}(e,\j)\g_{ab}+i\g^5A_\m\Big)\ve\equiv \big(\mathscr{D}_\m+i\g^5A_\m\big)\ve\,.
\eal
In these expressions $\o_\m{}^{ab}(e,\j)$ denotes the torsion-full spin connection
\be
\o_\m{}^{ab}(e,\j)\equiv\o_\m{}^{ab}(e)+\frac14\big(\lbar\j_a\g_\m\j_b+\lbar\j_\m\g_a\j_b-\lbar\j_\m\g_b\j_a\big)\,,
\ee
where $\o_\m{}^{ab}(e)$ is the torsion-free metric compatible connection.

The quantum effective action \eqref{W} of any $\cn=1$ SCFT is invariant under the local symmetry transformations \eqref{sugra-trans}, up to local expressions in the background supergravity fields that comprise the multiplet of superconformal anomalies. In particular, there exists a renormalization scheme such that, under an infinitesimal local symmetry transformation with parameters $\Omega=(\xi,\s,\l,\theta,\ve,\h)$, the quantum effective action transforms as
\be\label{W-trans}
\d_\Omega\mathcal W[e,A,\psi]=\int d^4x\, e\big(\s\mathcal A_W-\theta\mathcal A_R-\bar\ve\mathcal A_Q+\bar\h\mathcal A_S\big)\,,
\ee
where the superconformal anomalies $\ca_W$, $\ca_R$, $\ca_Q$ and $\ca_S$ take the form \cite{Papadimitriou:2019gel}
\bal\label{anomalies}
\ca_W=&\;\frac{c}{16\p^2}\Big(W^2-\frac{8}{3}F^2\Big)-\frac{a}{16\p^2} E+\co(\j^2)\,,\NO\\
\ca_R=&\;\frac{(5a-3c)}{27\p^2}\;\wt F F+\frac{(c-a)}{24\p^2}\cp\,,\NO\\
\ca_Q=&\;-\frac{(5a-3c)i}{9\p^2}\wt F^{\m\n}A_\m\g^5\f_\n+\frac{(a-c)}{6\p^2}\nabla_\m\big(A_\r \wt R^{\r\s\m\n} \big)\g_{(\n}\j_{\s)}-\frac{(a-c)}{24\p^2}F_{\m\n} \wt R^{\m\n\r\s} \g_\r\j_\s+\co(\j^3)\,,\NO\\
\ca_{S}=&\;\frac{(5a-3c)}{6\p^2}\wt F^{\m\n}\Big(\cd_\m-\frac{2i}{3}A_\m\g^5\Big)\j_{\n}+\frac{ic}{6\p^2} F^{\m\n}\big(\g_{\m}{}^{[\s}\d_{\n}^{\r]}-\d_{\m}^{[\s}\d_{\n}^{\r]}\big)\g^5\cd_\r\j_\s\\
&+\frac{3(2a-c)}{4\p^2}P_{\m\n}g^{\m[\n}\g^{\r\s]}\cd_\r\j_\s+\frac{(a-c)}{8\p^2}\Big(R^{\m\n\r\s}\g_{\m\n}-\frac12Rg_{\m\n}g^{\m[\n}\g^{\r\s]}\Big)\cd_\r\j_\s+\co(\j^3)\,.\NO
\eal

The anomaly coefficients $a$ and $c$ in these expressions depend on the specific SCFT and are normalized such that for $N_\c$ free chiral and $N_v$ free vector multiplets \cite{Anselmi:1997am}
\be
a=\frac{1}{48}(N_\c+9N_v)\,,\qquad c=\frac{1}{24}(N_\c+3N_v)\,.
\ee
Recall that $E$ and  $W^2$ in \eqref{anomalies} denote the Euler density and  the square of the Weyl tensor respectively,  while  $\cp$ is the Pontryagin density. Namely,  
\bal
W^2\equiv&\; W_{\m\n\r\s}W^{\m\n\r\s}=R_{\m\n\r\s}R^{\m\n\r\s}-2R_{\m\n}R^{\m\n}+\frac13R^2\,,\NO\\
E=&\;R_{\m\n\r\s}R^{\m\n\r\s}-4R_{\m\n}R^{\m\n}+R^2\,,\NO\\
\cp\equiv&\;\frac12\e^{\k\l\m\n}R_{\k\l\r\s}R_{\m\n}{}^{\r\s}=\wt R^{\m\n\r\s}R_{\m\n\r\s}\,,
\eal
where 
\be
\label{dualR}
\wt R_{\m\n\r\s}\equiv\frac12\e_{\m\n}{}^{\k\l}R_{\k\l\r\s}\,.
\ee
In addition, $P_{\m\n}$ in \eqref{anomalies} denotes the Schouten tensor 
\be\label{Schouten}
P_{\m\n}=\frac12\Big(R_{\m\n}-\frac16Rg_{\m\n}\Big)\,,
\ee
and we have defined 
\be
F^2\equiv F_{\m\n}F^{\m\n}\,,\qquad F\wt F\equiv \frac12 \e^{\m\n\r\s}F_{\m\n}F_{\r\s}\,,\qquad \wt F_{\m\n}\equiv\frac12 \e_{\m\n}{}^{\r\s}F_{\r\s}\,.
\ee

The anomalous transformation \eqref{W-trans} of the quantum effective action, together with the definition of the currents \eqref{consistent-currents}, leads to the superconformal Ward identities \cite{Papadimitriou:2019gel}  
\bbxd
\bal
\label{WardIDs}
&e^a_\m\nabla_\n\<\ct^\n_a\>+\nabla_\n(\lbar\j_\m  \<\cs^\n\>)-\lbar\j_\n\overleftarrow \cd_\m \<\cs^\n\>-F_{\m\n}\<\cj^\n\>\NO\\
&\hspace{2.cm}+A_\m\big(\nabla_\n\<\cj^\n\>+i\lbar\j_\n \g^5\<\cs^\n\>\big)-\o_\m{}^{ab}\Big(e_{\n [a}\<\ct^\n_{b]}\>+\frac14\lbar\j_\n\g_{ab}\<\cs^\n\>\Big)=0\,,\NO\\
&e^a_\m\<\ct^\m_a\>+\frac12\lbar\j_\m \<\cs^\m\>=\ca_W\,,\NO\\
&e_{\m[a} \<\ct^\m_{b]}\>+\frac14\lbar\j_\m\g_{ab} \<\cs^\m\>=0\,,\NO\\
&\nabla_\m \<\cj^\m\>+i\lbar\j_\m\g^5 \<\cs^\m\>=\ca_R\,,\NO\\
&\cd_\m \<\cs^\m\>-\frac12\g^a\j_\m \<\ct^\m_a\>-\frac{3i}{4}\g^5\f_\m\<\cj^\m\>=\ca_Q\,,\NO\\
&\g_\m \<\cs^\m\>-\frac{3i}{4}\g^5\j_\m\<\cj^\m\>=\ca_{S}\,.
\eal
\ebxd
The transformation \eqref{W-trans} of the quantum effective action, and hence the local terms in the Ward identities \eqref{WardIDs}, may be modified by adding local terms to the effective action $\cw[e,A,\j]$. For example, adding the term 
\be\label{mixedCT}
\int A\wedge\Tr\Big(\G\wedge d\G+\frac23 \G\wedge\G\wedge\G\Big)\,,
\ee
with $\G^\m{}_{\n}\equiv\G^{\m}_{\n\r}dx^\r$ the Christoffel connection, breaks the diffeomorphism invariance of the effective action and modifies the form of its anomalous transformation under the rest of the local symmetries. In particular, the addition of this term with a specific coefficient eliminates the Pontryagin term from the R-symmetry anomaly $\ca_R$, as is reviewed e.g. in \cite{Jensen:2012kj}. Although the fermionic anomalies $\ca_Q$ and $\ca_S$ can be modified by the addition of such local terms, there exists no local term that sets them to zero within conformal supergravity. It is these fermionic anomalies $\ca_Q$ and $\ca_S$ that determine the supersymmetric Casimir energy \cite{Papadimitriou:2017kzw}.

\section{Bardeen-Zumino terms and covariant currents}
\label{sec:BZ}

The anomalies \eqref{anomalies} are solutions of the Wess-Zumino (WZ) consistency conditions for the local symmetry transformations \eqref{sugra-trans} of $\cn=1$ conformal supergravity \cite{Papadimitriou:2019gel}. As such, they correspond to the so called {\em consistent anomalies}, while the operators defined in \eqref{consistent-currents} are known as the {\em consistent currents}. As a consequence of the R-symmetry anomaly, the consistent currents are not gauge invariant, while if the local term \eqref{mixedCT} is added to the effective action, the consistent currents are not diffeomorphism covariant either.    

The gauge and diffeomorphism covariance of the R-current $\cj^\m$ and of the stress tensor $\ct^\m_a$ can be restored by adding local Bardeen-Zumino (BZ) terms to the currents \cite{Bardeen:1984pm}. These terms are not related to the choice of renormalization scheme discussed above, since they cannot be expressed as derivatives of a local term in the effective action. Instead, they arise from a Chern-Simons action in five dimensions that cancels the R-symmetry/mixed anomaly through the mechanism of anomaly inflow (see e.g. \cite{Jensen:2012kj} for an extensive discussion). 

However, the BZ terms for the R-current and the energy-momentum tensor are already encoded in the form of the anomalies \eqref{anomalies} \cite{Papadimitriou:2017kzw,Papadimitriou:2019gel}. This observation was understood in \cite{Minasian:2021png} as a direct consequence of the WZ consistency conditions. As an example, let us consider the WZ condition\footnote{We adopt the convention of \cite{Papadimitriou:2019gel}, where the transformations act on the transformation parameters as well. This is analogous to the BRST treatment of the WZ consistency conditions, where the transformation parameters are replaced by ghost fields that themselves transform. The same conclusions are reached in the convention that the transformation parameters do not transform -- see section 3.2 of \cite{Minasian:2021png}.}
\be\label{WZdiffgauge}
(\d_\x\d_\th-\d_\th\d_\x)\cw=0\,.
\ee
In the absence of the local term \eqref{mixedCT}, $\d_\x\cw=0$ and so $\d_\x\d_\th\cw=0$ and $\d_\th\d_\x\cw=0$ separately. These imply respectively that $\d_\th\cw$ is diffeomorphism invariant and $\d_\x\cw$ is gauge invariant. Focusing on $\d_\x\cw$, it can be expressed in terms of the consistent currents \eqref{consistent-currents} as \cite{Papadimitriou:2019gel}
\be\label{diffeo-Ward}
\d_\x\cw=-\int d^4x\,e\,\x^\m\big(e^a_\m\nabla_\n\<\ct^\n_a\>+\nabla_\n(\lbar\j_\m  \<\cs^\n\>)-\lbar\j_\n\overleftarrow \cd_\m \<\cs^\n\>-F_{\m\n}\<\cj^\n\>+A_\m\ca_R\big)\,,
\ee
whose bosonic part is not manifestly gauge invariant. As we will now demonstrate, the form of the R-symmetry anomaly $\ca_R$ is such that this expression can be written in manifestly gauge invariant form in terms of the {\em covariant} R-current and energy-momentum tensor. 

Writing the R-symmetry anomaly explicitly, the bosonic terms in \eqref{diffeo-Ward} take the form
\bal
&\;e^a_\m\nabla_\n\<\ct^\n_a\>-F_{\m\n}\<\cj^\n\>+A_\m\ca_R\NO\\\
=&\;e^a_\m\nabla_\n\<\ct^\n_a\>-F_{\m\n}\<\cj^\n\>+\frac{(5a-3c)}{54\p^2}\; \e^{\k\l\r\s}F_{\k\l}F_{\r\s}A_\m+\frac{(c-a)}{48\p^2}\e^{\k\l\r\s}R_{\k\l\a\b}R_{\r\s}{}^{\a\b} A_\m\,.
\eal
Antisymmetrizing five indices in four dimensions gives zero. Hence, 
$F_{[\m\k}F_{\l\r}A_{\s]}=0$ and $R_{[\k\l\a\b}R_{\r\s}{}^{\a\b} A_{\m]}=0$, which imply respectively the identities
\be\label{SchoutenFlip-F}
\e^{\k\l\r\s}F_{\k\l}F_{\r\s}A_\m=-4\e^{\k\l\r\s}F_{\m\k}F_{\l\r}A_\s\,,
\ee
and
\be\label{SchoutenFlip-R}
\e^{\k\l\r\s}R_{\k\l\a\b}R_{\r\s}{}^{\a\b}A_\m=-4\e^{\k\l\r\s}R_{\m\k\a\b}R_{\l\r}{}^{\a\b}A_\s\,.
\ee

Therefore, the bosonic part of \eqref{diffeo-Ward} becomes  
\bal
&\;e^a_\m\nabla_\n\<\ct^\n_a\>-F_{\m\n}\<\cj^\n\>+A_\m\ca_R\NO\\\
=&\;e^a_\m\nabla_\n\<\ct^\n_a\>-F_{\m\n}\<\cj^\n_{\rm cov}\>-\frac{(c-a)}{12\p^2}\e^{\k\l\r\s}R_{\m\k\a\b}R_{\l\r}{}^{\a\b}A_\s\,,
\eal
where the covariant R-current is given by
\be
\<\cj^\m_{\rm cov}\>=\<\cj^\m\>+P_{BZ}^\m\,,
\ee
with the Bardeen-Zumino term
\bbxd
\vskip.4cm
\be\label{BZ-R-current}
P_{BZ}^\m=\frac{2(5a-3c)}{27\p^2}\; \e^{\m\l\r\s}F_{\l\r}A_\s\,.
\ee
\ebxd
Since the local term \eqref{mixedCT} is absent, the consistent R-current $\cj^\m$ transforms as tensor under diffeomorphisms and so the relevant BZ term need only restore gauge invariance. 

The BZ term for the energy-momentum tensor $\ct^{\m\n}=\ct^{\m a}e^\n_a$ takes the form \cite{Jensen:2012kj}
\bal
P^{\m\n}_{BZ}=&\;-\frac12\nabla_\l\big(X^{\l\m\n}+X^{\l\n\m}-X^{\m\n\l}\big)\,,\NO\\
X^{\m\l}{}_\n=&\;-\frac{(c-a)}{12\p^2}\big(\e^{\m\r\k\s}R^\l{}_{\n\k\s}+\e^{\l\r\k\s}R^\m{}_{\n\k\s}\big)A_\r\,.
\eal
With a bit of algebra this can be simplified to
\bbxd
\vskip.4cm
\be\label{BZ-stress-tensor}
P^{\m\n}_{BZ}=\frac{(c-a)}{6\p^2}\nabla_\l\big(\e^{\a\r\k\s}R^{\l\b}{}_{\k\s}A_\rho\big)\d^{\m}_{(\a}\d^{\n}_{\b)}\,.
\ee
\ebxd
Evaluating its covariant divergence we find
\bal
\nabla_\m P_{BZ}^{\m\n}=&\;\frac{(c-a)}{12\p^2}\nabla_\m\nabla_\l\big(\e^{\m\r\k\s}R^{\l\n}{}_{\k\s}A_\r+\e^{\n\r\k\s}R^{\l\m}{}_{\k\s}A_\r\big)\\
=&\;\frac{(c-a)}{12\p^2}\Big(\big([\nabla_\m,\nabla_\l]+\nabla_\l\nabla_\m\big)(\e^{\m\r\k\s}R^{\l\n}{}_{\k\s}A_\r)+\frac12[\nabla_\m,\nabla_\l](\e^{\n\r\k\s}R^{\l\m}{}_{\k\s}A_\r)\Big)\NO\\
=&\;\frac{(c-a)}{24\p^2}\Big([\nabla_\m,\nabla_\l](2\e^{\m\r\k\s}R^{\l\n}{}_{\k\s}A_\r-\e^{\n\r\k\s}R^{\m\l}{}_{\k\s}A_\r)+\nabla_\l(\e^{\m\r\k\s}R^{\l\n}{}_{\k\s}F_{\m\r})\Big)\,,\NO
\eal
where we made use of the second Bianchi identity $\nabla_{[\m}R^{\k\l}{}_{\n\r]}=0$.

Moreover, using the first Bianchi identity, $R^{\k}{}_{[\l\n\r]}=0$, we obtain
\be
[\nabla_\m,\nabla_\l](2\e^{\m\r\k\s}R^{\l\n}{}_{\k\s}A_\r-\e^{\n\r\k\s}R^{\m\l}{}_{\k\s}A_\r)=2\e^{\r\a\k\s}R^\n{}_{\a\m\l}R^{\m\l}{}_{\k\s}A_\r\,,
\ee
and hence
\be
\nabla_\m P_{BZ}^{\m\n}=\frac{(c-a)}{12\p^2}\e^{\r\a\k\s}R^\n{}_{\a\m\l}R^{\m\l}{}_{\k\s}A_\r+\nabla_\m L^{\m\n}=\frac{(c-a)}{24\p^2}\cp A^\n+\nabla_\m L^{\m\n}\,,
\ee
where we have used \eqref{SchoutenFlip-R} and
\bbxd
\vskip.4cm 
\be\label{L}
L^{\m\n}=\frac{(c-a)}{12\p^2}F_{\r\s}\wt R^{\r\s\m\n}\,.
\ee
\ebxd
It follows that the bosonic part of \eqref{diffeo-Ward} takes the form  
\bbxd
\vskip.2cm
\be
e^a_\m\nabla_\n\<\ct^\n_a\>-F_{\m\n}\<\cj^\n\>+A_\m\ca_R
=\nabla_\n\<\ct^\n_{\text{cov}\,\m}\>-F_{\m\n}\<\cj^\n_{\rm cov}\>-\nabla_\n L^\n{}_\m\,,
\ee
\ebxd
where
\be
\<\ct_{\text{cov}}^{\m\n}\>=\<\ct^{\m\n}\>+P^{\m\n}_{BZ}\,.
\ee
Since $L_{\m\n}$ is gauge invariant, we have demonstrated that all terms in \eqref{diffeo-Ward} can be made manifestly gauge invariant in terms of the covariant R-current and energy-momentum tensor.

The covariant R-current and energy-momentum tensor help simplify not only the diffeomorphism and R-symmetry Ward identities, but also those for $\cq$- and $\cs$-supersymmetry. In particular, in terms of the covariant currents, the Ward identities \eqref{WardIDs} take the form
\bbxd
\bal
\label{covWardIDs}
&e^a_\m\nabla_\n\<\ct^\n_{\text{cov}\,a}\>+\nabla_\n(\lbar\j_\m  \<\cs^\n\>)-\lbar\j_\n\overleftarrow \cd_\m \<\cs^\n\>-F_{\m\n}\<\cj_{\text{cov}}^\n\>=\ca_{D\,\m}^{\rm cov}\,,\NO\\
&e^a_\m\<\ct^\m_{\text{cov}\,a}\>+\frac12\lbar\j_\m \<\cs^\m\>=\ca^{\rm cov}_W\,,\NO\\
&e_{\m[a} \<\ct^\m_{\text{cov}\,b]}\>+\frac14\lbar\j_\m\g_{ab} \<\cs^\m\>=0\,,\NO\\
&\nabla_\m \<\cj_{\rm cov}^\m\>+i\lbar\j_\m\g^5 \<\cs^\m\>=\ca_R^{\rm cov}\,,\NO\\
&\cd_\m \<\cs^\m\>-\frac12\g^a\j_\m \<\ct^\m_{a\,\text{cov}}\>-\frac{3i}{4}\g^5\f_\m\<\cj^\m_{\rm cov}\>=\ca_Q^{\rm cov}\,,\NO\\
&\g_\m \<\cs^\m\>-\frac{3i}{4}\g^5\j_\m\<\cj^\m_{\rm cov}\>=\ca_{S}^{\rm cov}\,.
\eal
\ebxd
with the covariant anomalies given by the simpler expressions
\bbxd
\bal\label{covanomalies}
\ca_{D\,\m}^{\rm cov}=&\;\nabla_\n L^\n{}_\m=\frac{(a-c)}{12\p^2}\nabla^\n (F_{\r\s}\wt R^{\r\s}{}_{\m\n})\,,\NO\\
\ca^{\rm cov}_W=&\;\ca_W=\frac{c}{16\p^2}\Big(W^2-\frac{8}{3}F^2\Big)-\frac{a}{16\p^2} E+\co(\j^2)\,,\NO\\
\ca_R^{\rm cov}=&\;\frac{(5a-3c)}{9\p^2}\;\wt F F+\frac{(c-a)}{24\p^2}\cp\,,\NO\\
\ca_Q^{\rm cov}=&\;\frac{(c-a)}{24\p^2}F_{\m\n} \wt R^{\m\n\r\s} \g_\r\j_\s+\co(\j^3)\,,\NO\\
\ca_{S}^{\rm cov}=&\;\Big[\frac{(5a-3c)}{6\p^2}\wt F^{\r\s}+\frac{ic}{6\p^2} F^{\m\n}\big(\g_{\m}{}^{[\s}\d_{\n}^{\r]}-\d_{\m}^{[\s}\d_{\n}^{\r]}\big)\g^5+\frac{3(2a-c)}{4\p^2}P_{\m\n}g^{\m[\n}\g^{\r\s]}\NO\\
&+\frac{(a-c)}{8\p^2}\Big(R^{\m\n\r\s}\g_{\m\n}-\frac12Rg_{\m\n}g^{\m[\n}\g^{\r\s]}\Big)\Big]\cd_\r\j_\s+\co(\j^3)\,.
\eal
\ebxd

\section{Conserved charges}
\label{sec:charges}

The superconformal Ward identities determine the conserved charges and their algebra. In preparation for the derivation of the supersymmetric Casimir energy in the subsequent sections, we consider the R-charge and the charges associated with conformal Killing vectors and spinors on supergravity backgrounds that admit Killing spinors. A key property of such backgrounds is that the superconformal anomalies are numerically zero, resulting in continuous families of conserved bosonic charges.

Starting with the R-charge, we define the one-parameter family of R-currents \cite{Papadimitriou:2017kzw}
\bbxd
\vskip.2cm
\be
\<\cj^\m_{\o_\cj}\>\equiv \<\cj^\m\>+\o_{\cj} P_{BZ}^\m\,,
\ee
\ebxd
where $P_{BZ}^\m$ is the BZ term given in \eqref{BZ-R-current}. Clearly, $\<\cj^\m_0\>$ is the consistent R-current, while $\<\cj^\m_1\>$ is the covariant one. On a bosonic background the divergence of this current is
\be
\nabla_\m\<\cj^\m_{\o_\cj}\>=(1+2\o_{\cj})\frac{5a-3c}{27\p^2}\wt FF+\frac{c-a}{24\p^2}\cp\,,
\ee
which is nonzero on a generic supergravity background. However, for conformal supergravity backgrounds that admit Killing spinors both terms in the R-symmetry anomaly are numerically zero, i.e. $\wt FF=0$ and $\cp=0$. As a result, on supersymmetric backgrounds there exists a continuous family of conserved R-charges defined as
\bbxd
\vskip.4cm
\be\label{R-charge}
Q_R^{\o_\cj}=\int_\cc d\s_\m \<\cj^\m_{\o_\cj}\>\,,
\ee
\ebxd
where $\cc$ is a Cauchy surface and $d\s_\m$ is the corresponding infinitesimal area element. Notice that although the anomalies are numerically zero on supersymmetric backgrounds, the BZ terms are not necessarily vanishing, and hence the value of the R-charge \eqref{R-charge} in general depends on the parameter $\o_\cj$. 

Let us now consider the conserved charges associated with conformal Killing vectors of supersymmetric backgrounds. A conformal Killing vector $\ck^\m$ satisfies the relations
\be\label{confKilling}
\mathcal L_\ck\, g_{\m\n}=\nabla_\m\ck_\n+\nabla_\n\ck_\m=\frac{1}{2}(\nabla_\r\ck^\r)g_{\m\n}\,,\qquad\mathcal L_\ck A_\m=\del_\m \L_\ck\,,
\ee
where $\cl_\ck$ denotes the Lie derivative with respect to $\ck^\m$ and $\L_\ck$ is an arbitrary R-symmetry gauge parameter. The Killing condition on the R-symmetry gauge field $A_\m$ is equivalent to the gauge-invariant condition $\cl_\ck F_{\m\n}=0$.  

As with the R-current, we define the continuous family of energy-momentum tensors
\bbxd
\vskip.2cm
\be
\<\ct^{\m\n}_{\o_\ct}\>\equiv \<\ct^{\m\n}\>+\o_{\ct} P_{BZ}^{\m\n}\,,
\ee
\ebxd
where $P_{BZ}^{\m\n}$ is the BZ term \eqref{BZ-stress-tensor}. Notice again that $\<\ct^{\m\n}_0\>$ corresponds to the consistent energy-momentum tensor, while $\<\ct^{\m\n}_1\>$ is the covariant one. The divergence of this current on a bosonic background is given by 
\be
\nabla_\m\<\ct^{\m}_{\o_\ct\,\n}\>=F_{\n\m}\<\cj^\m_{\rm cov}\>+\nabla_\m\big((\o_\ct-1)P^\m_{BZ\,\n}+L^\m{}_\n\big)\,.
\ee

In order to determine the conserved charge associated with a conformal Killing vector $\ck^\m$ we evaluate the divergence
\bal
&\;\nabla_\m\big((\<\ct^{\m}_{\o_\ct\,\n}\>+A_\n\<\cj^\m_{\o_\cj}\>)\ck^\n\big)\NO\\
=&\;\nabla_\m\<\ct^{\m}_{\o_\ct\,\n}\>\ck^\n+\frac14\<\ct^{\m}_{\o_\ct\,\m}\>\nabla_\r\ck^\r+\pa_\m (A_\n \ck^\n)\<\cj^\m_{\o_\cj}\>+A_\n\ck^\n\nabla_\m\<\cj^\m_{\o_\cj}\>\NO\\
=&\;\ck^\n F_{\n\m}\<\cj^\m_{\rm cov}\>+(F_{\m\n} \ck^\n+\pa_\m\L_\ck)\<\cj^\m_{\o_\cj}\>\NO\\
&\;+\ck^\n\nabla_\m\big((\o_\ct-1)P^\m_{BZ\,\n}+L^\m{}_\n\big)+\frac14\ca_W\nabla_\r\ck^\r+A_\n\ck^\n\nabla_\m\<\cj^\m_{\o_\cj}\>\NO\\
=&\;(\o_{\cj}-1)F_{\m\n} \ck^\n P_{BZ}^\m+\nabla_\m(\L_\ck\<\cj^\m_{\o_\cj}\>)\NO\\
&\;+\ck^\n\nabla_\m\big((\o_\ct-1)P^\m_{BZ\,\n}+L^\m{}_\n\big)+\frac14\ca_W\nabla_\r\ck^\r+(A_\n\ck^\n-\L_\ck)\nabla_\m\<\cj^\m_{\o_\cj}\>\,.
\eal
The relation \eqref{SchoutenFlip-F} implies that 
\be
F_{\m\n} \ck^\n P_{BZ}^\m=\frac{(5a-3c)}{27\p^2}F\wt F A_\m\ck^\m\,,
\ee
and so we conclude that
\bal
&\;\nabla_\m\big(\<\ct^{\m}_{\o_\ct\,\n}\>\ck^\n+(A_\n\ck^\n-\L_\ck)\<\cj^\m_{\o_\cj}\>\big)
=\frac14\ca_W\nabla_\r\ck^\r+\o_\ct\ck^\n\nabla_\m L^\m{}_\n\NO\\
&\;+\big(3\o_{\cj}A_\n\ck^\n-(1+2\o_{\cj})\L_\ck\big)\frac{5a-3c}{27\p^2} F\wt F+(\o_\ct A_\n\ck^\n-\L_\ck)\frac{c-a}{24\p^2}\cp\,.
\eal

On all the backgrounds we consider the quantities $\wt FF$, $\cp$, as well as $\nabla_\m L^{\m\n}$ vanish numerically, while the conformal anomaly, $\ca_W$, is not necessarily zero \cite{Cassani:2013dba}. However, the product $\ca_W\nabla_\r\ck^\r$ is vanishing for all cases we are interested in, either due to $\ca_W$ being zero, or due to $\ck^\m$ being Killing instead of conformal Killing. It follows that for all such backgrounds there exists a two-parameter family of conserved currents associated with the Killing vector (or conformal Killing if $\ca_W=0$) $\ck^\m$, namely
\be
\nabla_\m\big(\<\ct^{\m}_{\o_\ct\,\n}\>\ck^\n+(A_\n\ck^\n-\L_\ck)\<\cj^\m_{\o_\cj}\>\big)=0\,,
\ee
and hence a two-parameter family of conserved charges
\bbxd
\vskip.4cm
\be\label{K-charge}
Q^{\o_\ct,\o_\cj}[\ck]=\int_\cc d\s_\m \big(\<\ct^{\m}_{\o_\ct\,\n}\>\ck^\n+(A_\n\ck^\n-\L_\ck)\<\cj^\m_{\o_\cj}\>\big)\,.
\ee
\ebxd

Finally, we consider the conserved charges associated with (conformal) Killing spinors of $\cn=1$ conformal supergravity. These are solutions of the Killing spinor equation obtained 
by setting the local symmetry transformation of the gravitino in \eqref{sugra-trans} to zero. On a bosonic background this leads to the Killing spinor equation 
\be\label{KSE-eta}
\cd_\m\ve_0=\g_\m\h_0\,.
\ee
In the following we will also need the Majorana conjugate equation
\be\label{KSE-eta-bar}
\bar\ve_0\overleftarrow \cd_\m=-\bar\h_0\g_\m\,.
\ee
Expressing $\h_0$ in terms of $\ve_0$ through the algebraic relation
\be
\h_0=\frac14\g^\n\cd_\n\ve_0\,,
\ee
the Killing spinor equation can be written in the form
\bbxd
\vskip.4cm
\be\label{KSE}
\cd_\m\ve_0=\frac14\g_\m\g^\n\cd_\n\ve_0\,.
\ee
\ebxd

On a bosonic background the Ward identities \eqref{WardIDs} imply that the supercurrent is covariantly conserved and has zero $\g$-trace, i.e.
\be
\cd_\m \<\cs^\m\>=0\,,\qquad \g_\m \<\cs^\m\>=0\,.
\ee
These in turn imply that\footnote{Note that the definition \eqref{consistent-currents} implies that the R-charge of the supercurrent $\<\cs^\m\>$ is minus that of the gravitino, and hence of the Killing spinor $\ve_0$.}
\bal
\nabla_\m(\bar\ve_0\<\cs^\m\>)=\bar\ve_0\overleftarrow \cd_\m\<\cs^\m\>=-\bar\h_0\g_\m\<\cs^\m\>=0\,,
\eal
and, hence, the quantity 
\bbxd
\vskip.4cm
\be\label{Q-charge}
Q[\ve_0]=\int_\cc d\s_\m \bar\ve_0\<\cs^\m\>\,,
\ee
\ebxd
corresponds to the conserved supercharge associated with the (conformal) Killing spinor $\ve_0$. Note that the Killing spinor $\ve_0$ must be commuting in order for the supercharge \eqref{Q-charge} to be Grassmann-valued.

\section{Supersymmetric Casimir energy}
\label{sec:SCE}

We are now in a position to obtain the general form of the supersymmetric Casimir energy on $\cn=1$ conformal supergravity backgrounds, generalizing the result of \cite{Papadimitriou:2017kzw} to generic SCFTs with $a\neq c$. Our starting point is the anomalous transformation of the supercurrent $\cs^\m$ under local $\cq$- and $\cs$-supersymmetries. These transformations follow directly from the anomalous superconformal Ward identities \eqref{WardIDs} and are given by \cite{Papadimitriou:2019gel}
\bbxd 
\bal\label{supercurrent-trans}
\d_\ve\<\cs^\m\>
=&\;\frac{1}{2}\g^a\ve\<\ct^\m_{\text{cov}\,a}\>+\frac{i}{8}\big(4\d^{[\m}_\n\d^{\r]}_\s+i\g^5 \e^\m{}_{\n}{}^{\r}{}_\s\big) \g^\n\g^5\cd_\r\big(\ve\<\cj^\s_{\text{cov}}\>\big)+\frac{(a-c)}{24\p^2}F_{\r\s}\wt R^{\r\s\m\n}\g_\n\ve\,,\NO\\
\rule{0cm}{1.0cm}\d_\h\<\cs^\m\>
=&\;\frac{3i}{4}\g^5\h\<\cj^\m_{\text{cov}}\>+\frac{(5a-3c)}{6\p^2}\cd_\n(\wt F^{\m\n}\h)-\frac{ic}{6\p^2}\big(\g^{[\m}{}_\r\d^{\n]}_\s-\d^{[\m}_\r\d^{\n]}_\s\big)\g^5\cd_\n(F^{\r\s}\h)\\
&-\frac{3(2a-c)}{4\p^2}\cd_\n\big(P_{\r\s}g^{\r[\s}\g^{\m\n]}\h\big)-\frac{(a-c)}{8\p^2}\cd_\n\Big[\Big(R^{\m\n\r\s}\g_{\r\s}-\frac12Rg_{\r\s}g^{\r[\s}\g^{\m\n]}\Big)\h\Big]\,.\NO
\eal
\ebxd

\subsection{Supercharge algebra on curved backgrounds}

Taking the transformation parameters $\ve$, $\h$ to be the (commuting) components of a conformal Killing spinor, $\ve_0$, $\h_0=\frac14\g^\n\cd_\n\ve_0$, the transformations \eqref{supercurrent-trans} compute the algebra of the corresponding supercharges through the relation    
\be\label{algebra}
\<\{Q[\ve_0],Q[\ve_0]\}\>=\int_\cc d\s_\m \,\bar\ve_0(\d_{\ve_0}+\d_{\eta_0})\<\cs^\m\>\,.
\ee
Since the anticommutator of the supercharges vanishes on BPS states, this relation determines the BPS relation among the bosonic conserved charges on an arbitrary supersymmetric background, and hence the general form of the supersymmetric Casimir energy. 

Our task, therefore, is to evaluate the r.h.s. of eq.~\eqref{algebra} using the supercurrent transformations \eqref{supercurrent-trans}. The term $\bar\ve_0\d_{\ve_0}\<\cs^\m\>$ takes the form
\be
\frac{1}{2}\bar\ve_0\g_\n\ve_0\Big(\<\ct^{\m\n}_{\text{cov}}\>+\frac{(a-c)}{24\p^2}F_{\r\s}\wt R^{\r\s\m\n}\Big)+\frac{i}{8}\bar\ve_0\big(4\d^{[\m}_\n\d^{\r]}_\s+i\g^5 \e^\m{}_{\n}{}^{\r}{}_\s\big) \g^\n\g^5\cd_\r\big(\ve_0\<\cj^\s_{\text{cov}}\>\big)\,.
\ee 
Integrating by parts, the term proportional to the covariant R-current becomes 
\be\label{JFA}
\nabla_\r\cv_{\cq}^{\m\r}-\frac{i}{8}\bar\ve_0\overleftarrow\cd_\r\big(4\d^{[\m}_\n\d^{\r]}_\s+i\g^5 \e^\m{}_{\n}{}^{\r}{}_\s\big) \g^\n\g^5\ve_0\<\cj^\s_{\text{cov}}\>\,,
\ee
where 
\be
\cv_{\cq}^{\m\r}=\frac{i}{8}\bar\ve_0\big(4\d^{[\m}_\n\d^{\r]}_\s+i\g^5 \e^\m{}_{\n}{}^{\r}{}_\s\big) \g^\n\g^5\ve_0\<\cj^\s_{\text{cov}}\>\,.
\ee
The Killing spinor equation \eqref{KSE-eta-bar} and the identity \eqref{gddg5e} in Appendix \ref{appendix:identities}, allow us to simplify the second term in \eqref{JFA} to 
\be
-\frac{3i}{4}(\bar\h_0\g^5\ve_0)\<\cj^\m_{\text{cov}}\>\,.
\ee
Therefore, we conclude that the term $\bar\ve_0\d_{\ve_0}\<\cs^\m\>$ takes the form
\bbxd
\be
\label{CasimirQ}
\bar\ve_0\d_{\ve_0}\<\cs^\m\>=\frac{1}{2}\bar\ve_0\g_\n\ve_0\Big(\<\ct^{\m\n}_{\text{cov}}\>+\frac{(a-c)}{24\p^2}F_{\r\s}\wt R^{\r\s\m\n}\Big)-\frac{3i}{4}(\bar\h_0\g^5\ve_0)\<\cj^\m_{\text{cov}}\>+\nabla_\r\cV_\cq^{\r\m}\,.
\ee
\ebxd

Let us next focus on the term
\bal\label{d-eta-S}
\bar\ve_0\d_{\h_0}\<\cs^\m\>
=&\;\frac{3i}{4}\bar\ve_0\g^5\h_0\<\cj^\m_{\text{cov}}\>+\frac{(5a-3c)}{6\p^2}\bar\ve_0\cd_\n(\wt F^{\m\n}\h_0)-\frac{ic}{6\p^2}\bar\ve_0\big(\g^{[\m}{}_\r\d^{\n]}_\s-\d^{[\m}_\r\d^{\n]}_\s\big)\g^5\cd_\n(F^{\r\s}\h_0)\NO\\
&\hspace{-1.5cm}-\frac{3(2a-c)}{4\p^2}\bar\ve_0\cd_\n\big(P_{\r\s}g^{\r[\s}\g^{\m\n]}\h_0\big)-\frac{(a-c)}{8\p^2}\bar\ve_0\cd_\n\Big[\Big(R^{\m\n\r\s}\g_{\r\s}-\frac12Rg_{\r\s}g^{\r[\s}\g^{\m\n]}\Big)\h_0\Big]\,,
\eal
and consider in turn all local terms. 

\begin{flushleft}
\hrule\hrule
$\frac{(5a-3c)}{6\p^2}\bar\ve_0\cd_\r(\wt F^{\m\r}\h_0):$
\vskip4pt\hrule\hrule
\end{flushleft}

Integrating by parts and using the Killing spinor equation \eqref{KSE-eta-bar} this term becomes 
\be\label{Ftilde1}
\nabla_\r\cv_{\cs_1}^{\m\r}+\frac{(5a-3c)}{6\p^2}\wt F^{\m\r} \bar\h_0\g_\r\h_0\,, 
\ee
where
\be
\cv_{\cs_1}^{\m\r}=\frac{(5a-3c)}{6\p^2}\wt F^{\m\r}\bar\ve_0\h_0\,.
\ee

\begin{flushleft}
\hrule\hrule
$-\frac{ic}{6\p^2}\bar\ve_0(\g^{[\m}{}_\r\d^{\n]}_\s-\d^{[\m}_\r\d^{\n]}_\s)\g^5\cd_\n(F^{\r\s}\h_0):$
\vskip4pt\hrule\hrule
\end{flushleft}

Similarly, this term can be written in the form
\be\label{gg5F}
\nabla_\r \cv^{\m\r}_{\cs_2} -\frac{ic}{6\p^2}\bar\h_0\g_\r\big(\g^{[\m}{}_\n\d^{\r]}_\s-\d^{[\m}_\n\d^{\r]}_\s\big)\g^5\h_0\,F^{\n\s}\,,
\ee
with
\be
\cv^{\m\r}_{\cs_2}=-\frac{ic}{6\p^2}\bar\ve_0\big(\g^{[\m}{}_\n\d^{\r]}_\s-\d^{[\m}_\n\d^{\r]}_\s\big)\g^5\h_0\,F^{\n\s}\,.
\ee

Moreover, the second term in \eqref{gg5F} gives
\bal\label{gg5F1}
-\frac{ic}{6\p^2}\bar\h_0\g_\r\big(\g^{[\m}{}_\n\d^{\r]}_\s-\d^{[\m}_\n\d^{\r]}_\s\big)\g^5\h_0\,F^{\n\s}&=-\frac{ic}{6\p^2}\bar\eta_0(\g_\s \g^{\m}{}_\n-\d^\m_\n\g_\s)\g^5\h_0 F^{\n\s}\NO\\
&=-\frac{ic}{6\p^2}\big(\bar\eta_0\g^\s \g^{\m\n}\g^5 \h_0\,F_{\n\s}-\bar\h_0\g_\s\g^5\h_0 F^{\m\s}\big)\,.
\eal
The second term vanishes, since $\bar\h_0\g_\s\g^5\h_0=0$ for commuting spinors, while 
\be\label{gsgmng5}
\bar\eta_0\g^\s \g^{\m\n}\g^5 \h_0=\bar\eta_0 \g^{\m\n}\g^\s\g^5 \h_0=\bar\eta_0 (\g^{\m\n\s}+\g^\m g^{\n\s}-\g^\n g^{\m\s})\g^5 \h_0=i\e^{\m\n\s\r}\bar\h_0\g_\r\h_0\,.
\ee

Hence, 
\be
-\frac{ic}{6\p^2}\bar\h_0\g_\r\big(\g^{[\m}{}_\n\d^{\r]}_\s-\d^{[\m}_\n\d^{\r]}_\s\big)\g^5\h_0\,F^{\n\s}=
\frac{c}{6\p^2}\e^{\m\n\s\r}F_{\n\s}\bar\h_0\g_\r\h_0=\frac{c}{3\p^2}\wt F^{\m\r}\,\bar\h_0\g_\r\h_0\,,
\ee
and, therefore, \eqref{gg5F} can be simplified to 
\be\label{tildeF2}
\nabla_\r\cv^{\m\r}_{\cs_2}+\frac{2c}{6\p^2}\wt F^{\m\r}\,\bar\h_0\g_\r\h_0\,.
\ee

\begin{flushleft}
\hrule\hrule
$-\frac{3(2a-c)}{4\p^2}\bar\ve_0\cd_\n(P_{\r\s}g^{\r[\s}\g^{\m\n]}\h_0):$
\vskip4pt\hrule\hrule
\end{flushleft}

Using integration by parts and the Killing spinor equation \eqref{KSE-eta-bar} this term becomes
\be\label{Pns}
\nabla_\r\cv_{\cs_3}^{\m\r}-\frac{3(2a-c)}{4\p^2} P_{\n\s}\,\bar\h_0\g_\r \,g^{\n[\s}\g^{\m\r]}\h_0\,,
\ee
where
\be
\cv_{\cs_3}^{\m\r}=-\frac{3(2a-c)}{4\p^2}\bar\ve_0(P_{\n\s}g^{\n[\s}\g^{\m\r]}\h_0)\,.
\ee

The second term in \eqref{Pns} can be simplified as
\bal
-\frac{3(2a-c)}{4\p^2} P_{\n\s}\,\bar\h_0\g_\r \,g^{\n[\s}\g^{\m\r]}\h_0
&=-\frac{3(2a-c)}{4\p^2}\frac23P_{\n\s}(-g^{\n\s}\bar\h_0\g^\m\h_0+g^{\m\n}\bar\h_0\g^\s\h_0)\NO\\
&=-\frac{(2a-c)}{2\p^2}\left(-\frac16R\,\d^\m_\r+P^\m_{\,\,\r}\right)\bar\h_0\g^\r\h_0\,,
\eal
where in the last step we used that $P\equiv P_{\n\s}g^{\n\s}=R/6$. Observing that 
\be
-\frac16R\,\d^\m{}_\r+P^\m{}_{\r}=\frac12\Big(R^\m{}_{\r}-\frac12 R\,\d^\m{}_\r\Big)\,,
\ee
we arrive at the final form of the expression \eqref{Pns}:
\be\label{Pnsfinal}
\nabla_\r\cv_{\cs_3}^{\m\r}-\,\frac{(2a-c)}{4\p^2}\Big(R^\m{}_{\r}-\frac12 R\,\d^\m{}_\r\Big)\bar\h_0\g^\r\h_0\,.
\ee

\begin{flushleft}
\hrule\hrule
$-\frac{(a-c)}{8\p^2}\bar\ve_0\,\cd_\n\big[\big(R^{\m\n\r\s}\g_{\r\s}-\frac12Rg_{\r\s}g^{\r[\s}\g^{\m\n]}\big)\h_0\big]:$
\vskip4pt\hrule\hrule
\end{flushleft}

This term is similarly written as
\be\label{gR}
\nabla_\r\cv_{\cs_4}^{\m\r}-\frac{(a-c)}{8\p^2}\bar\h_0\g_\r\Big(R^{\m\r\n\s}\g_{\n\s}-\frac12Rg_{\n\s}g^{\n[\s}\g^{\m\r]}\Big)\h_0\,,
\ee
with
\be
\cv_{\cs_4}^{\m\r}=-\frac{(a-c)}{8\p^2}\bar\ve_0\Big(R^{\m\r\n\s}\g_{\n\s}-\frac12Rg_{\n\s}g^{\n[\s}\g^{\m\r]}\Big)\h_0\,.
\ee

The first term in the square bracket in \eqref{gR} gives 
\bal
\bar\h_0\g_\r(R^{\m\r\n\s}\g_{\n\s})\h_0&=-R^{\m}_{\,\,\,\,\r\n\s}\,\bar\h_0(\g^{\n\s\r}+\g^\n g^{\s\r}-\g^\s g^{\n\r})\h_0\NO\\
&=-2R^\m_{\,\,\,\r}\,\bar\h_0\g^\r\h_0\,,
\eal
where we used the property $\bar\h_0\g_\r\g_{\n\s}\h_0=-\bar\h_0\g_{\n\s}\g_\r\h_0$ for commuting spinors. Moreover, the second term in the square bracket in \eqref{gR} becomes
\be
\frac12R\,g_{\n\s}(\bar\h_0\g_\r g^{\n[\s}\g^{\m\r]}\h_0)=\frac12R\,g_{\n\s}\left(-\frac23\,g^{\n\s}\bar\h_0\g^\m\h_0+\frac23\,g^{\m\n}\bar\h_0\g^\s\h_0\right)=- R \,\bar\h_0\g^\m\h_0\,,
\ee
where we used the identity \eqref{gammagidentity} in Appendix \ref{appendix:identities} with $\ve=\h$. 

We therefore conclude that
\be
-\frac{(a-c)}{8\p^2}\bar\h_0\g_\r\Big(R^{\m\r\n\s}\g_{\n\s}-\frac12Rg_{\n\s}g^{\n[\s}\g^{\m\r]}\Big)\h_0=\frac{(a-c)}{4\p^2}\left(R^\m_{\,\,\,\r}-\frac{1}{2}R\,\d^\m_\r\right)\bar\h_0\g^\r\h_0\,,
\ee
and hence \eqref{gR} reduces to
\be\label{gR2}
\nabla_\r\cv_{\cs_4}^{\m\r}+\,\frac{(a-c)}{4\p^2}\left(R^\m_{\,\,\,\r}-\frac{1}{2}R\,\d^\m_\r\right)\bar\h_0\g^\r\h_0\,.
\ee

Returning to the overall variation \eqref{d-eta-S}, we observe that \eqref{Ftilde1} and \eqref{tildeF2} combine to 
\be
\frac{(5a-c)}{6\p^2}\wt F^{\m\r} \bar\h_0\g_\r\h_0\,,
\ee
while \eqref{Pnsfinal} and \eqref{gR2} reduce to 
\be
-\frac{a}{4\p^2}\Big(R^\m_{\,\,\,\r}-\frac{1}{2}R\,\d^\m_\r\Big)\bar\h_0\g^\r\h_0\,.
\ee
Therefore, gathering all terms, the variation \eqref{d-eta-S} is written as
\bbxd
\be\label{CasimirS}
\bar\ve_0\d_{\h_0}\<\cs^\m\>=\frac{3i}{4}\bar\ve_0\g^5\h_0\<\cj^\m_{\text{cov}}\>+\Big[\frac{(5a-c)}{6\p^2}\wt F^{\m}{}_\r-\frac{a}{4\p^2}\Big(R^\m_{\,\,\,\r}-\frac{1}{2}R\,\d^\m_\r\Big)\Big]\bar\h_0\g^\r\h_0+\nabla_\r \mathcal V_\cs^{\m\r}\,,
\ee
\ebxd
with
\bal
\cv_\cs^{\m\r}=&\;\frac{(5a-3c)}{6\p^2}\bar\ve_0(\wt F^{\m\r}\h_0)+\frac{ic}{6\p^2}\ve_0\big(\g^{[\m}{}_\n\d^{\r]}_\s-\d^{[\m}_\n\d^{\r]}_\s\big)\g^5(F^{\n\s}\h_0)\NO\\
&\;-\frac{3(2a-c)}{4\p^2}\bar\ve_0\big(P_{\n\s}g^{\n[\s}\g^{\m\r]}\h\big)-\frac{(a-c)}{8\p^2}\ve_0\Big(R^{\m\r\n\s}\g_{\n\s}-\frac12Rg_{\n\s}g^{\n[\s}\g^{\m\r]}\Big)\h_0\,.
\eal

Combining \eqref{CasimirQ} and \eqref{CasimirS}, and using the commuting spinor identity $\bar\h_0\g^5\ve_0=-\bar\ve_0\g^5\h_0$, we arrive at the result 
\bbxd
\bal
\label{CasimirQS}
\bar\ve_0(\d_{\ve_0}+\d_{\eta_0})\<\cs^\m\>=&\;\frac{1}{2}\bar\ve_0\g_\n\ve_0\Big(\<\ct^{\m\n}_{\text{cov}}\>+\frac{(a-c)}{24\p^2}F_{\r\s}\wt R^{\r\s\m\n}\Big)+\frac{3i}{2}(\bar\ve_0\g^5\h_0)\<\cj^\m_{\text{cov}}\>\\
&\;+\Big[\frac{(5a-c)}{6\p^2}\wt F^{\m}{}_\r-\frac{a}{4\p^2}\Big(R^\m_{\,\,\,\r}-\frac{1}{2}R\,\d^\m_\r\Big)\Big]\bar\h_0\g^\r\h_0+\nabla_\r (\cv_\cq^{\r\m}+\cv_\cs^{\m\r})\,.\NO
\eal
\ebxd
Inserting this in \eqref{algebra} and dropping the total derivative terms we conclude that BPS states satisfy the integral constraint
\bbxd
\vskip-.1cm
\bal\label{preBPS}
0=&\;\int_\cc d\s_\m\, \bar\ve_0(\d_{\ve_0}+\d_{\eta_0})\<\cs^\m\>\NO\\
=&\;\int_\cc d\s_\m\Big\{\frac{1}{2}\bar\ve_0\g_\n\ve_0\Big(\<\ct^{\m\n}_{\text{cov}}\>+\frac{(a-c)}{24\p^2}F_{\r\s}\wt R^{\r\s\m\n}\Big)+\frac{3i}{2}\bar\ve_0\g^5\h_0\,\<\cj^\m_{\text{cov}}\>\NO\\
&\;\hspace{4.5cm}+\Big[\frac{(5a-c)}{6\p^2}\wt F^{\m}{}_\r-\frac{a}{4\p^2}\Big(R^\m_{\,\,\,\r}-\frac{1}{2}R\,\d^\m{}_\r\Big)\Big]\bar\h_0\g^\r\h_0\Big\}\,.
\eal
\ebxd
We will now show that this constraint corresponds to the BPS relation among the bosonic charges of supersymmetric states and determines the supersymmetric Casimir energy.

\subsection{BPS relation and the supersymmetric Casimir energy}

In order to relate the constraint \eqref{preBPS} to the conserved R-charge and conformal Killing charges, respectively \eqref{R-charge} and \eqref{K-charge}, we begin by showing that the spinor bilinear $\bar\ve_0\g^\m\ve_0$ is a conformal Killing vector. In fact, for later use we show more generally that if $\ve_0$ and $\ve_0'$ are conformal Killing spinors, then the spinor bilinear  
\be
\ck^\m(\ve_0,\ve_0')\equiv \bar\ve_0'\g^\m\ve_0\,,
\ee 
is a conformal Killing vector. Thus, we compute
\bal
\nabla_\m\ck_{\n}(\ve_0,\ve_0')=&\;\bar\ve_0'\overleftarrow\cd_\m\g_\n\ve_0+\bar\ve_0'\g_\n\cd_\m\ve_0\NO\\
=&\;-\bar\h_0'\g_\m\g_\n\ve_0+\bar\ve_0'\g_\n\g_\m\h_0\NO\\
=&\;\bar\ve_0\g_\n\g_\m\h_0'+\bar\ve_0'\g_\n\g_\m\h_0\,,
\eal
where we used the Killing spinor equations \eqref{KSE-eta} and \eqref{KSE-eta-bar} and the fact that for commuting spinors $\bar\h_0'\g_\m\g_\n\ve_0=-\bar\ve_0\g_\n\g_\m\h_0'$. Hence,
\be
\nabla_\m\ck_{\n}(\ve_0,\ve_0')+\nabla_\n\ck_{\m}(\ve_0,\ve_0')=2g_{\m\n}(\bar\ve_0\h_0'+\bar\ve_0'\h_0)=\frac12g_{\m\n}\nabla_\r\ck^\r(\ve_0,\ve_0')\,,
\ee
which confirms that $\ck^\m(\ve_0,\ve_0')$ is a conformal Killing vector. 

It follows that 
\be\label{K0}
\ck_0^\m\equiv \bar\ve_0\g^\m\ve_0\,,
\ee
is a conformal Killing vector. In fact, for the supersymmetric backgrounds with a conserved R-charge, i.e. numerically zero R-symmetry anomaly, such as those we consider below, this Killing spinor bilinear is further constrained to be a Killing vector \cite{Cassani:2012ri}. As a result, \eqref{preBPS} can be written in the form
\bal
0=&\;Q^{\o_\ct,\o_\cj}[\ck_0]+\int_\cc d\s_\m\Big\{\big(3i(\bar\ve_0\g^5\h_0)-A_\n\ck_0^\n+\L_{\ck_0}\big)\<\cj^\m_{\o_\cj}\>\NO\\
&+\Big((1-\o_\ct)P_{BZ}^{\m\n}+\frac{(a-c)}{24\p^2}F_{\r\s}\wt R^{\r\s\m\n}\Big)\ck_{0\n}+3i(\bar\ve_0\g^5\h_0)(1-\o_\cj)P_{BZ}^\m\NO\\
&\;+\Big[\frac{(5a-c)}{3\p^2}\wt F^{\m}{}_\n-\frac{a}{2\p^2}\Big(R^\m_{\,\,\,\n}-\frac{1}{2}R\,\d^\m_\n\Big)\Big]\bar\h_0\g^\n\h_0\Big\}\,.
\eal

However, the coefficient multiplying the R-current $\<\cj^\m_{\o_\cj}\>$ is a constant, since
\be
\pa_\m\big(3i(\bar\ve_0\g^5\h_0)-A_\n\ck_0^\n+\L_{\ck_0}\big)=3i\pa_\m(\bar\ve_0\g^5\h_0)-F_{\m\n} \ck_0^\n\,,
\ee
and 
\bal
3i\pa_\m(\bar\ve_0\g^5\h_0)=&\;3i\bar\ve_0\overleftarrow \cd_\m\g^5\h_0+3i\bar\ve_0\g^5\cd_\m\h_0\NO\\
=&\;-3i\bar\h_0\g_\m\g^5\h_0-\frac{3i}{2}\bar\ve_0\g^5\Big(P_{\m\n}+\frac{2i}{3}F_{\m\n}\g^5-\frac{1}{3}\wt F_{\m\n}\Big)\g^\n\ve_0\NO\\
=&\;F_{\m\n} \ck_0^\n\,,
\eal
where we used the fact that $\bar\h_0\g_\m\g^5\h_0=\bar\ve_0\g_\m\g^5\ve_0=0$ for commuting spinors, and \cite{Papadimitriou:2019gel}
\be
\cd_\m\h_0=-\frac12\Big(P_{\m\n}+\frac{2i}{3}F_{\m\n}\g^5-\frac{1}{3}\wt F_{\m\n}\Big)\g^\n\ve_0\,.
\ee

We conclude that \eqref{preBPS} can be further simplified to
\bbxd
\vskip.25cm
\be\label{genBPS}
Q^{\o_\ct,\o_\cj}[\ck_0]+\F_{\ck_0}Q_R^{\o_\cj}+Q_{\rm local}^{\o_\ct,\o_\cj}[\ck_0]=0\,,
\ee
\ebxd
where

\bbxd
\vskip.2cm
\be\label{R-potential}
\F_{\ck_0}\equiv 3i(\bar\ve_0\g^5\h_0)-A_\n\ck_0^\n+\L_{\ck_0}=\text{const.}\,,
\ee
\ebxd
is the R-charge chemical potential and
\bbxd
\bal\label{locCharge}
Q_{\rm local}^{\o_\ct,\o_\cj}[\ck_0]\equiv &\;\int_\cc d\s_\m\Big\{\Big((1-\o_\ct)P_{BZ}^{\m\n}+\frac{(a-c)}{24\p^2}F_{\r\s}\wt R^{\r\s\m\n}\Big)\ck_{0\n}\\
&\;\hskip-2.cm+(1-\o_\cj)(\F_{\ck_0}+A_\n\ck_0^\n-\L_{\ck_0})P_{BZ}^\m+\Big[\frac{(5a-c)}{3\p^2}\wt F^{\m}{}_\n-\frac{a}{2\p^2}\Big(R^\m_{\,\,\,\n}-\frac{1}{2}R\,\d^\m_\n\Big)\Big]\bar\h_0\g^\n\h_0\Big\}\,,\NO
\eal
\ebxd
is a {\em local} charge that depends only on the supergravity background. As we demonstrate in the next section, when applied to a suitable globally defined Killing vector, the relation \eqref{genBPS} determines the energy of a BPS state in terms of the conserved R-charge, with \eqref{locCharge} corresponding to the supersymmetric Casimir energy.

\section{Casimir energy on $\bb R\times S^3$}
\label{sec:sphere}

The BPS relation \eqref{genBPS} and the local charge \eqref{locCharge} hold for any supersymmetric background of $\cn=1$ conformal supergravity with numerically vanishing superconformal anomalies. Such backgrounds have been studied extensively \cite{Samtleben:2012gy,Klare:2012gn,Dumitrescu:2012ha,Liu:2012bi,Dumitrescu:2012at,Kehagias:2012fh,Closset:2012ru,Samtleben:2012ua,Cassani:2012ri,deMedeiros:2012sb,Hristov:2013spa} (see also \cite{Blau:2000xg,Kuzenko:2012vd} for earlier work). In this section we will focus on a concrete background for which the supersymmetric Casimir energy has been computed in the literature by other means. In particular, we will apply our general result to an example in the class of backgrounds with topology $\bb R\times S^3$ (or $S^1\times S^3$ in Euclidean signature) that admit a single Majorana supercharge.

\subsection{Supersymmetric backgrounds with $\bb R\times S^3$ topology}

Following \cite{Assel:2014paa}, we consider four-dimensional backgrounds of topology $\bb R\times S^3$ that admit a non-singular complete direct product metric of the form\footnote{The corresponding Euclidean backgrounds are obtained by setting $t = i\t$, see e.g. \cite{Cassani:2012ri,Cassani:2013dba}.}  
\bbxd
\vskip.2cm
\be\label{Lmetric}
ds^2=-\Omega^2(\r)dt^2+f^2(\r)d\r^2+m_{IJ}(\r)\,d\varphi_I\,d\varphi_J,\quad I,J=1,2\,,
\ee
\ebxd
where $\vf_I\in[0,2\p]$, $\r\in[0,1]$, while the functions $\O(\r)$, $f(\r)$ are positive definite, as is the symmetric matrix $m_{IJ}(\r)$. This metric possesses an $\bb R\times U(1)^2$ isometry ($U(1)^3$ in its Euclidean form) corresponding to the commuting Killing vectors $\pa_t$, $\pa_{\vf_1}$ and $\pa_{\vf_2}$.

Demanding that the background admits a Majorana supercharge requires the existence of a globally defined null Killing vector of the form
\bbxd
\vskip.4cm
\be\label{SKV}
\ck=\frac12(\del_t+b_I\del_{\vf_I})=\frac12(\del_t+b_1\del_{\vf_1}+b_2\del_{\vf_2})\,,
\ee
\ebxd
where $b_1$, $b_2$ are real parameters. The requirement that this is null fixes 
\be\label{Omega^2}
\O^2=b^Im_{IJ}b^J \,.
\ee
Notice that the metric on $S^3$, which is parameterized by $f(\r)$ and $m_{IJ}(\r)$, is not constrained by supersymmetry.   

Finally, the globally well defined R-symmetry gauge field takes the form \cite{Assel:2014paa}\footnote{This differs by an overall minus sign compared to \cite{Assel:2014paa} due to different conventions.}
\bbxd 
\vskip.4cm
\be\label{Lgauge-field}
A=-\frac{\O}{8fc}\pa_\r\Big[(c^2+a_\c^2) \Big(\frac{d \vf_1}{b_1} - \frac{d \vf_2}{b_2}\Big)+2a_\c \Big(\frac{d\vf_1}{b_1}+\frac{d\vf_2}{b_2}+dt\Big)\Big]-\frac{1}{2} d\o\,,
\ee
\ebxd
where
\be
a_\chi=\frac{1}{\O^2}(b_1^2m_{11}-b_2^2m_{22})\,, \quad c=\frac{2|b_1b_2|}{\O^2}\sqrt{\det(m_{IJ})}\,,\quad \o = {\rm sgn}(b_1) \vf_1 + {\rm sgn}(b_2) \vf_2\,.
\ee

These expressions describe a family of non-singular backgrounds that admit a single Majorana supercharge. They are parameterized by the arbitrary non-singular metric on $S^3$ and the real parameters $b_1$, $b_2$. The squashed (Berger) three-sphere is a special case of these backgrounds \cite{Assel:2014paa}. However, for simplicity we will illustrate our results by considering the background with arbitrary $b_1$, $b_2$ and the round metric on $S^3$.

\subsection{Round $S^3$ with arbitrary $b_1,b_2$}

Defining the angular coordinate $\th=\p\r$, the metric \eqref{Lmetric} corresponding to the round $S^3$ is 
\be\label{S3-mentric}
ds^2=-\Omega^2dt^2+d\theta^2+m_{IJ}d\varphi_Id\varphi_J\,,
\ee
with
\bbxd
\vskip.4cm
\be\label{mIJ-round-sphere}
m_{11} = 4 \cos^2\frac{\theta}{2} \,,\qquad  m_{22} = 4 \sin^2\frac{\theta}{2} \,,\qquad m_{12}=0\,,\qquad f=\p\,.
\ee
\ebxd

These expressions for $m_{IJ}$ completely determine the background. In particular, 
\be
\O^2= b^Im_{IJ}b^J= 2\big(b_1^2+b_2^2+(b_1^2-b_2^2)\cos\th\big)\,,
\ee
\be
a_\chi=\frac{1}{\O^2}(b_1^2m_{11}-b_2^2m_{22})=\frac{b_1^2-b_2^2+(b_1^2+b_2^2)\cos\th}{b_1^2+b_2^2+(b_1^2-b_2^2)\cos\th}\,,
\ee
and
\be
c=\frac{2|b_1b_2|}{\O^2}\sqrt{\det(m_{IJ})}=\frac{2|b_1b_2|\sin\th}{b_1^2+b_2^2+(b_1^2-b_2^2)\cos\th}\,.
\ee
Inserting these in \eqref{Lgauge-field} we obtain the R-symmetry gauge field
\bbxd
\vskip.4cm
\be\label{A-S3}
A = \frac{{\rm sgn}(b_1b_2)}{\O}(b_2d\vf_1+b_1d\vf_2)+\frac{|b_1b_2|}{\Omega}\,dt  -\frac{1}{2}d\o \, .
\ee
\ebxd

\subsection*{Killing spinors}

The Killing spinors are solutions of the conformal Killing equation \eqref{KSE}. In order to solve this equation we introduce a suitable local frame for the background metric \eqref{S3-mentric}, namely 
\bbxd
\bal\label{veilbeins}
e^0=&\;\O dt\,,\NO\\
e^1=&\;\frac{\O}{2b_1b_2}\big((1+a_\c)b_2d\vf_1+(1-a_\c)b_1d\vf_2\big)\,,\NO\\
e^2=&\;d\th\,,\NO\\
e^3=&\;\frac{\O}{2b_1b_2}(b_2d\vf_1-b_1d\vf_2)\,,
\eal
\ebxd
so that 
\be
ds^2=-(e^0)^2+(e^1)^2+(e^2)^2+(e^3)^2\,.
\ee

We also use the Weyl representation of the gamma matrices so that $\g^\m=e^\m_a\g^a$ with
\be
\g^a=\left(\begin{array}{cc}
0 &\s^a\\
\bar\s^a& 0\\
\end{array}\right),\qquad \s_a=(-\bb 1,\s_i),\quad \bar\s_a=(\bb 1,\s_i)\,,
\ee
and
\be
\g^{ab}=\left(\begin{array}{cc}
\s^{ab} & 0\\
0 & \bar\s^{ab}\\
\end{array}\right),\qquad \s_{ab}=\frac12[\s_a,\bar\s_b],\quad \bar\s_{ab}=\frac12[\bar\s_a,\s_b]\,.
\ee
Moreover, the chirality matrix takes the form
\be
\g^5=\left(\begin{array}{cc}
\bb 1 & 0\\
0 & -\bb1\\
\end{array}\right)\,.
\ee

The commuting Majorana spinors $\ve_0$ and $\h_0$ can then be expressed as 
\be
\ve_0=\left(\begin{array}{c}
\e \\
\tilde\e\\
\end{array}\right)\,, \qquad 
\h_0=\left(\begin{array}{c}
\zeta \\
\tilde\zeta\\
\end{array}\right)\,,
\ee
where $\e$, $\z$ are left-handed two-component Weyl spinors and $\tilde\e\equiv i\s_2\e^*$, $\tilde\zeta\equiv i\s_2\zeta^*$. Therefore, the Killing spinor equation \eqref{KSE} is equivalent to the two-component spinor equations 
\bbxd
\vskip.4cm
\be\label{KSE-2comp}
\cd^L_\m\e=\s_\m \tilde\zeta\,,\quad \tilde\zeta=\frac14\bar\s^\m\,\cd^L_\m\e\,,\qquad \cd^R_\m\tilde\e=\bar\s_\m \zeta\,\,,\quad \zeta=\frac14\s^\m\,\cd^R_\m\e\,,
\ee
\ebxd
where the chiral derivatives are $\cd^L_\m=\del_\m+\frac14\omega^{ab}_\m\s_{ab}+iA_\m$ and  $\cd^R_\m=\del_\m+\frac14\omega^{ab}_\m\bar\s_{ab}-iA_\m$.  

For generic $b_1$ and $b_2$ the Killing spinor equation \eqref{KSE} admits the unique solution 
\bbxd
\vskip.4cm
\be
\ve_0=e^{\frac i2\o}\sqrt{\frac{\O}{2}}\left(\begin{matrix} -1\\ 1 \\ 0 \\ 0 \end{matrix}\right)+e^{-\frac i2\o}\sqrt{\frac{\O}{2}}\left(\begin{matrix} 0\\ 0 \\ 1 \\ 1 \end{matrix}\right)\,,
\ee
\ebxd
while $\h_0=\frac14\g^\n\cd_\n\ve_0$ takes the form
\bbxd
\vskip.4cm
\be
\h_0=-\frac{i}{(2\O)^{3/2}}e^{\frac i2\o}\left(\begin{matrix} 0\\ 0 \\ -2|b_1b_2|+\O\O' \\ 2|b_1b_2|+\O\O' \end{matrix}\right)-\frac{i}{(2\O)^{3/2}}e^{-\frac i2\o}\left(\begin{matrix} 2|b_1b_2|+\O\O'\\ 2|b_1b_2|-\O\O' \\ 0 \\ 0 \end{matrix}\right)\,.
\ee
\ebxd 

Having determined the conformal Killing spinor, we can evaluate the three spinor bilinears that enter in the BPS relation \eqref{genBPS} and the local charge \eqref{locCharge}. We find 
\bbxd
\bal\label{bilinears}
&\bar\ve_0\g^\m\ve_0=-4\ck^\m\,,\NO\\
&3i\bar\ve_0\g^5\h_0-A_\m \bar\ve_0\g^\m\ve_0=-(|b_1|+|b_2|)\,,\NO\\
&\bar\h_0\g^\m\h_0=\frac{1}{4}\ck^\m-\frac{1}{2\O^2}\big(b_1^2+b_2^2,0,(b_1^2-b_2^2)b_1,-(b_1^2-b_2^2)b_2\big)\,,
\eal
\ebxd
where $\ck^\m$ is the globally defined null Killing vector \eqref{SKV}. These expressions reaffirm the results we obtained earlier. In particular, $\bar\ve_0\g^\m\ve_0$ is a Killing vector, as it should, while the R-charge potential \eqref{R-potential} is indeed constant (note that $\L_{\ck_{0}}=0$ in this case since $\cl_{\ck_0}A_\m=0$).

\subsection*{Casimir energy}

We now have all ingredients in order to evaluate the local charge \eqref{locCharge} on this supersymmetric background. Since $\ck_0^\m\equiv\bar\ve_0\g^\m\ve_0=-4\ck^\m\sim -2\pa_t$, the Casimir energy is given by
\be
\ce_{\rm Casimir}^{\o_\ct,\o_\cj}\equiv -\frac{1}{2}Q_{\rm local}^{\o_\ct,\o_\cj}[\ck_0]=Q_{\rm local}^{\o_\ct,\o_\cj}[2\ck]\,,
\ee
and, therefore, it takes the form
\bbxd
\bal\label{ECasimir}
\ce_{\rm Casimir}^{\o_\ct,\o_\cj}\equiv &\;\int_\cc d\s_\m\Big\{\big(2(1-\o_\ct)P_{BZ}^{\m\n}-L^{\m\n}\big)\ck_\n-(1-\o_\cj)\frac{3i}{2}(\bar\ve_0\g^5\h_0)P_{BZ}^\m\NO\\
&\;\hspace{3.cm}+\Big[\frac{a}{4\p^2}\Big(R^\m_{\,\,\,\n}-\frac{1}{2}R\,\d^\m_\n\Big)-\frac{(5a-c)}{6\p^2}\wt F^{\m}{}_\n\Big]\bar\h_0\g^\n\h_0\Big\}\,,
\eal
\ebxd
where $\ck^\m$ is the globally defined Killing vector in \eqref{SKV}, $L^{\m\n}$ is given in \eqref{L}, while the BZ terms are given in \eqref{BZ-R-current} and \eqref{BZ-stress-tensor}.

Evaluating the R-current BZ term on the background specified by the metric \eqref{S3-mentric} and R-symmetry gauge field \eqref{A-S3} we find  
\be\label{BZ-R-current-S3}
P_{BZ}^\m=\frac{2(5a-3c)}{27\p^2}\frac{(b_1^2-b_2^2)}{2\O^4}\big(-{\rm sgn}(b_1b_2)(|b_1|-|b_2|),0,-|b_1|b_2,|b_2|b_1\big)\,.
\ee
Similarly, the only nonzero components of the stress tensor BZ term \eqref{BZ-stress-tensor} take the form
\bal\label{BZ-stress-tensor-S3}
P^{t\vf_1}_{BZ}=&\;\frac{(c-a)}{3\p^2}\frac{(b_1^2-b_2^2)}{\O^8}(2\O-5|b_1|)\,b_1^2b_2^2\,{\rm sgn}(b_2)\,,\NO\\
P^{t\vf_2}_{BZ}=&\;-\frac{(c-a)}{3\p^2}\frac{(b_1^2-b_2^2)}{\O^8}(2\O-5|b_2|)\,b_1^2b_2^2\,{\rm sgn}(b_1)\,.
\eal
Also, the nonzero components of the tensors $L^{\m\n}$  and $\wt F_{\m\n}$ are respectively
\bal\label{L-S3}
L^{t\vf_1}=&\;\frac{(a-c)}{24\p^2}\frac{(b_1^2-b_2^2)^2}{\O^6}|b_1|{\rm sgn}(b_2)(1-\cos\th)\,,\NO\\
L^{t\vf_2}=&\;\frac{(a-c)}{24\p^2}\frac{(b_1^2-b_2^2)^2}{\O^6}|b_2|{\rm sgn}(b_1)(1+\cos\th)\,,\NO\\
L^{\vf_1\vf_2}=&\;-\frac{(a-c)}{48\p^2}\frac{(b_1^2-b_2^2)}{\O^4}|b_1b_2|\,,
\eal
and
\bal
\wt F_{t\vf_1}=&\;-\frac{(b_1^2-b_2^2)}{\O^2}|b_1|{\rm sgn}(b_2)(1+\cos\th)\,,\NO\\
\wt F_{t\vf_2}=&\;\frac{(b_1^2-b_2^2)}{\O^2}|b_2|{\rm sgn}(b_1)(1-\cos\th)\,,\NO\\
\wt F_{\vf_1\vf_2}=&\;\frac{2(b_1^2-b_2^2)}{\O^4}|b_1b_2|\sin^2\th\,,
\eal
while the Einstein tensor on the supersymmetric background \eqref{S3-mentric} takes the form
\be
R_{\m\n}-\frac12 R g_{\m\n}={\rm diag}\Big(0,\frac{b_1^2+b_2^2}{\O^2},\frac{b_2^2m_{11}}{\O^2}\Big(\frac{4b_1^2}{\O^2}+1\Big),\frac{b_1^2m_{22}}{\O^2}\Big(\frac{4b_2^2}{\O^2}+1\Big)\Big)-\frac34 g_{\m\n}\,.
\ee

Putting everything together we can evaluate the integral \eqref{ECasimir} on the Cauchy surface defined by the constant time slices. The result is 
\bbxd
\bal\label{ECasimir-S3}
\ce_{\rm Casimir}^{\o_\ct,\o_\cj}
=&\;\frac{3a}{2}\frac{|b_1||b_2|}{(|b_1|+|b_2|)}+\frac{(|b_1|-|b_2|)^2}{9b_1b_2(|b_1|+|b_2|)}\big(c_1(|b_1|^2+|b_2|^2)+c_2|b_1||b_2|\big)\,,
\eal
\ebxd
where the coefficients $c_1$ and $c_2$ are given by
\bbxd
\bal
c_1=&\;(5a-3c)(\o_\cj-1)-(a-c)(2\o_\ct-1)+\frac32(5a-c)\,,\NO\\
c_2=&\;9a\,{\rm sgn}(b_1b_2)+2(5a-3c)(\o_\cj-1)-3(a-c)(\o_\ct-1)+\frac32(5a-c)\,.
\eal
\ebxd

For $|b_1|=|b_2|$, in which case the metric \eqref{S3-mentric} is conformally flat, \eqref{ECasimir} reduces to 
\bbxd
\vskip.4cm
\be\label{ECasimir-CF}
\ce_{\rm Casimir}=\frac{3a}{4}|b|\,,\qquad |b_1|=|b_2|=|b|\,,
\ee
\ebxd
independently of the values of $\o_\ct$ and $\o_\cj$. This result is in agreement with the expression \eqref{CasimirS0} in the scheme where the coefficient of the $R^2$ counterterm is set to zero.

\section{Discussion}
\label{sec:discussion}


In this work we have used the conformal current multiplet algebra obtained in \cite{Papadimitriou:2019gel} in order to determine the supersymmetric Casimir energy of generic four-dimensional $\cn=1$ superconformal field theories on conformal supergravity backgrounds admitting Killing spinors. In the simplest supersymmetric and fully diffeomorphism invariant scheme, our general result for the supersymmetric Casimir energy is given in eq.~\eqref{ECasimir}. This expression involves an integral over a Cauchy surface of a local density that depends only on the supersymmetric background. It is also a function of two real parameters, $\o_\cj$ and $\o_\ct$, which reflect the continuum of possible definitions of respectively the R-charge and conformal Killing charges on supersymmetric backgrounds. The choice $\o_\cj=0$, $\o_\ct=0$ corresponds to conserved charges defined in terms of the consistent R-current and energy-momentum tensor, while $\o_\cj=1$, $\o_\ct=1$ corresponds to charges defined in terms of the covariant currents. It is not surprising that this is the only choice resulting in a fully covariant and gauge invariant Casimir energy.

We illustrated our result for the supersymmetric Casimir energy by evaluating the general expression \eqref{ECasimir} on a class of supersymmetric backgrounds with topology $\bb R\times S^3$ that admit a single Majorana supercharge and depend on two real complex structure parameters, $b_1$, $b_2$. The resulting expression in given is eq.~\eqref{ECasimir-S3} and can be compared with the supersymmetric Casimir energy in eq.~\eqref{Esquashed}, obtained through supersymmetric localization \cite{Kim:2012ava,Assel:2014paa}. It is immediately clear that the two expressions for the supersymmetric Casimir energy do not coincide. The mismatch between our expression \eqref{ECasimir-S3} for the supersymmetric Casimir energy on $\bb R\times S^3$ and the localization result \eqref{Esquashed} is not surprising, however. Similar discrepancies have been observed when comparing the localization result with holographic computations of the Casimir energy \cite{BenettiGenolini:2016qwm}. In that special case, corresponding to $a=c$, the authors of \cite{BenettiGenolini:2016qwm} showed that the two expressions for the supersymmetric Casimir energy differ by a local counterterm that explicitly breaks the diffeomorphism invariance of the holographic effective action, i.e. of the bulk on-shell action. The same counterterm ensures that the holographic effective action depends on the background only via the transversely holomorphic foliation of $S^3$ \cite{BenettiGenolini:2016tsn}.  

In \cite{Papadimitriou:2017kzw} it was shown that the non-holomorphicity of the partition function is not an artifact of holographic renormalization, but rather a direct consequence of the fermionic superconformal Ward identities, which had not been properly incorporated in the earlier literature. In particular, the partition function for any $\cn=1$ SCFT with arbitrary anomaly coefficients $a$ and $c$ cannot be simultaneously holomorphic and fully diffeomorphism invariant. This is precisely the reason for the discrepancy between the two expressions \eqref{ECasimir-S3} and \eqref{Esquashed} for the supersymmetric Casimir energy. Our result \eqref{ECasimir-S3} corresponds to a diffeomorphism invariant and supersymmetric scheme, while the localization result \eqref{ECasimir-S3} corresponds to a holomorphic scheme that explicitly breaks diffeomorphism invariance. The complete form of the local counterterm that relates the two renormalization schemes for generic anomaly coefficients $a$ and $c$ and a detailed comparison of the two expressions for the supersymmetric Casimir energy will be reported elsewhere \cite{SusyCasHol}.

\section*{Acknowledgments}

The work of PP was supported  by the National Research Foundation of Korea (NRF) grant funded by the Korea government (MSIT) (No.~2023R1A2C1006975) and from the JRG Program at the APCTP through the Science and Technology Promotion Fund and Lottery Fund of the Korean Government. PP would like to thank the National and Kapodistrian University of Athens for the hospitality during the completion of this work. IP would like to thank the Isaac Newton Institute for Mathematical Sciences, Cambridge, for support and hospitality during the programme ``Black holes: bridges between number theory and holographic quantum information" where work on this paper was undertaken. This work was supported by EPSRC grant no EP/R014604/1.

\appendix

\renewcommand{\thesection}{\Alph{section}}
\renewcommand{\theequation}{\Alph{section}.\arabic{equation}}

\section{Conventions and spinor identities}
\label{appendix:identities}

Our spacetime and spinor conventions are those of \cite{Freedman:2012zz}. The tangent space metric is $\h=\diag(-1,1,1,1)$ and the Levi-Civita symbol $\ve_{\m\n\r\s}=\pm 1$ satisfies $\ve_{0123}=1$. Moreover, the Levi-Civita tensor is defined as $\e_{\m\n\r\s}=\sqrt{-g}\;\ve_{\m\n\r\s}=e\;\ve_{\m\n\r\s}$. Finally, the chirality matrix is
\be
\g^5=i\g_0\g_1\g_2\g_3\,.
\ee

Several gamma matrix and spinor identities we use in this manuscript are given in Appendix A of \cite{Papadimitriou:2019gel}. Here we quote only the gamma matrix identities
\be\label{eg-identities}
\g^{\m\n\r\s}=i\e^{\m\n\r\s}\g^5,\qquad \g^{\m\n\r}=i \e^{\m\n\r\s}\g_\s\g^5,\qquad 
\g^{\m\n}=\frac i2\e^{\m\n}{}_{\r\s}\g^{\r\s}\g^5\,,
\ee
and the {\em anticommuting} spinor flip relations under Majorana conjugation
\bal\label{anticommuting-flip}
&\bar\ve\,\g^5\h=\bar\h\g^5\ve\,,\NO\\
&\bar\ve\,\g_\m\h=-\bar\h\g_\m\ve\,,\NO\\
&\bar\ve\,\g_\m \g^5\h=\bar\h \,\g_\m\g^5\,,\NO\ve\\
&\ve\, \g^\s\g^{\m\n}\g^5\h=-\bar\h \,\g^{\m\n}\g^\s\g^5\ve\,,\NO\\
&\ve\, \g^\s\g^{\m\n}\h=\bar\h \,\g^{\m\n}\g^\s\ve\,.
\eal

However, the evaluation of the Casimir energy involves bilinears of {\em commuting} spinors, for which the flip relations in \eqref{anticommuting-flip} hold with an additional minus sign. Using these we now prove two identities for commuting spinors that we use extensively in our analysis. 

\bbxd
\vskip.3cm
\be\label{gddg5e}
\bar\h\g_\r\big(4\d^{[\m}_\n\d^{\r]}_\s+i\g^5 \e^\m{}_{\n}{}^{\r}{}_\s\big) \g^\n\g^5\ve
=-6\d^\m_\s\,\bar\h\g^5\ve\,.
\ee
\ebxd
\begin{center}\textbf{Proof}\end{center}
The first term can rewritten as follows:
\be\label{grddgng5}
\begin{split}
\bar\h\g_\r(4\d^{[\m}_\n\d^{\r]}_\s)\g^\n\g^5\ve&=2\,\bar\h\g_\r(\d^\m_\n\d^\r_\s-\d^\r_\n\d^\m_\s)\g^\n\g^5\ve\\
&=2\,\bar\h\g_\s\g^\m\g^5\ve-2\d^\m_\s\,\bar\h\g_\n \g^\n\g^5\ve\\
&=2g_{\k\s}\bar\h\g^\k\g^\m\g^5\ve-8\,\d^\m_\s\,\bar\h\g^5\ve\,,
\end{split}
\ee
where in the second step we used that $\g_\n\g^\n=4$. The first  term of eq.~\eqref{grddgng5} becomes
\be\label{gkgmg5}
\begin{split}
2g_{\k\s}\bar\h\g^\k\g^\m\g^5\ve&=2g_{\k\s}\bar\h(\g^{\k\m}+g^{\k\m})\g^5\ve\\
&=2g_{\k\s}\,\bar\h\g^{\k\m}\g^5\ve+2\d^\m_\s\,\bar\h\g^5\ve\,,
\end{split}\ee
and expressing  $\g^{\m\n}$ in terms of the Levi-Civita symbol using eq.~\eqref{eg-identities}, we get 
\be
\begin{split}
2g_{\k\s}\,\bar\h\g^{\k\m}\g^5\ve&=ig_{\k\s}\,\e^{\k\m}{}_{\n\r}\,\bar\h\g^{\n\r}\ve=-i \e^{\m}{}_{\s\n\r}\,\bar\h\g^{\n\r}\ve\,.
\end{split}
\ee
Inserting this in eq.~\eqref{gkgmg5} we obtain 
\be
2g_{\k\s}\bar\h\g^\k\g^\m\g^5\ve=-i \e^{\m}{}_{\s\n\r}\,\bar\h\g^{\n\r}\ve+2\d^\m_\s\,\bar\h\g^5\ve\,,
\ee
so that eq.~\eqref{grddgng5} is written as 
\be\label{gdeltagg}
\bar\h\g_\r(4\d^{[\m}_\n\d^{\r]}_\s)\g^\n\g^5\ve=-i \e^{\m}{}_{\s\n\r}\,\bar\h\g^{\n\r}\ve-6\d^\m_\s\,\bar\h\g^5\ve\,\,.
\ee
The second term on the l.h.s. of eq.~\eqref{gddg5e} is easily treated and gives
\be\label{gg5epsilon}
i\bar\h\g_\r\g^5 \e^\m{}_{\n}{}^{\r}{}_\s \g^\n\g^5\ve=i\e^{\m}{}_{\s\r\n}\,\bar\h\g^\r\g^\n\ve_0=i\e^{\m}{}_{\s\n\r}\,\bar\h\g^{\n\r}\ve_0\,\,.
\ee
Summing eq.~\eqref{gdeltagg} and eq.~\eqref{gg5epsilon} the Levi-Civita part cancels out and get eq.~\eqref{gddg5e}. \,\,\,\,\,$\blacksquare$ 

\bigskip

\bbxd
\vskip.4cm
\be
\label{gammagidentity}
 \bar\ve\g_\r \,g^{\n[\s}\g^{\m\r]}\h=-\frac{i}{3}\e^{\s\m\n\r}\,\bar\h\g_\r\g^5\ve-\frac23\,g^{\n\s}\bar\ve\g^\m\h+\frac23\,g^{\m\n}\bar\ve\g^\s\h\,.
\ee
\ebxd
\begin{center}\textbf{Proof}\end{center}

We  first manipulate the $\g$-term, namely
\be
\g_\r \,g^{\n[\s}\g^{\m\r]}=\frac13\g_\r\left(g^{\n\s}\g^{\m\r}+g^{\n\r}\g^{\s\m}+g^{\n\m}\g^{\r\s}\right)=\frac13(-3g^{\n\s}\g^\m+\g^\n\g^{\s\m}+3\,g^{\m\n}\g^\s)\,,
\ee
where we used that $\g_\r\g^{\r\m}=3\g^\m$. For commuting spinors $\bar\ve \g^{\n}\g^{\m\r}\h=-\bar\h \g^{\m\r}\g^\n\ve$ and hence
\be\label{grgmr}
\bar\ve\g_\r \,g^{\n[\s}\g^{\m\r]}\h=-g^{\n\s}\bar\ve\g^\m\h+g^{\m\n}\bar\ve\g^\s\h-\frac13 \,\bar\h\g^{\s\m}\g^\n\ve\,.
\ee
Note that the last term may be written as 
\be\label{gsmgn}
\bar\h\g^{\s\m}\g^\n\ve=\bar\h\left(\g^{\s\m\n}+\g^\s g^{\m\n}-\g^\m g^{\s\n}\right)\ve\,.
\ee
The first term in this expression gives $\bar\h \g^{\s\m\n}\ve=i\e^{\s\m\n\r}\,\bar\h\g_\r\g^5\ve$. Inserting  eq.~\eqref{gsmgn} in eq.~\eqref{grgmr} we obtain eq.~\eqref{gammagidentity}. \hspace{10.5cm}$\blacksquare$

Note that for $\ve=\h$, relation \eqref{gammagidentity} is simplified to
\be
\label{ggidentity}
\bar\h\g_\r\,g^{\n[\s}\g^{\m\r]}\h=-\frac23\,g^{\n\s}\bar\h\g^\m\h+\frac23\,g^{\m\n}\bar\h\g^\s\h\,.
\ee


\bibliographystyle{JHEP}
\bibliography{CScasimir}

\end{document}